\documentclass[]{aastex}

\usepackage{emulateapj5}
\usepackage{onecolfloat}
\usepackage{graphicx} 
\usepackage{ulem}
\usepackage{rotating}
\usepackage{lscape}
\usepackage{epsfig}
\usepackage{portland}

\newcommand{\kms}{km~s$^{-1}$ }
\newcommand{\cm}[1]{\, {\rm cm^{#1}}}

\newcommand{\sci}[1]{{\rm \; \times \; 10^{#1}}}

\newcommand{\Ha}{H$\alpha$}

\begin{document}

\twocolumn[
\title{Star Cluster Populations in the Outer Disks of Nearby Galaxies}
\author{St\'ephane Herbert-Fort$^1$, Dennis Zaritsky$^1$, John Moustakas$^2$,\\ 
Andrea Di Paola$^3$, Richard W.\ Pogge$^4$, Roberto Ragazzoni$^5$}
\vspace{0.15cm}
\affil{
$^1$University of Arizona/Steward Observatory, 933 N Cherry Avenue, Tucson, AZ 85721 USA\\
 (email: s.herbertfort@gmail.com, dennis.zaritsky@gmail.com) \\
$^2$Center for Astrophysics and Space Sciences, University of California, San Diego, La Jolla, CA 92093, USA\\
$^3$INAF, Osservatorio Astronomico di Roma, via di Frascati 33, I-00040 Monteporzio, Italy\\
$^4$Department of Astronomy, The Ohio State University, 140 W. 18th
Avenue, Columbus, OH 43210-1173\\
$^5$INAF, Osservatorio Astronomico di Padova, vicolo dell'Osservatorio 5, I-35122 Padova, Italy\\
}

\begin{abstract}

We present a Large Binocular Telescope (LBT) imaging study that characterizes the star cluster component of 
nearby galaxy outer disks (beyond the optical radius $R_{25}$).  Expanding
on the pilot project of \cite{HF09}, we
present deep ($\sim27.5$ mag $V$-band point-source limiting magnitude) $U$-\ and $V$-band 
imaging of six galaxies:  IC 4182, NGC 3351, NGC 4736, NGC 4826, NGC 5474, and NGC 6503.  
We find that the
outer disk of each galaxy is populated with marginally-resolved star clusters with masses $\sim10^3 M_{\sun}$ and ages up to $\sim1$ Gyr 
(masses and ages are limited by the depth of our imaging and uncertainties are large given how photometry can be strongly affected by the presence or absence of a few stars in such low mass systems), and that they are typically found
out to at least $2 R_{25}$ but sometimes as far as 3 to 4 $R_{25}$ -- even beyond the 
apparent \ion{H}{1} disk.  The mean rate of cluster formation for $1 R_{25} \le  R \le 1.5R_{25}$
is at least one every $\sim2.5$ Myr and the clusters are spatially correlated with the \ion{H}{1}, most strongly with higher density gas near the periphery of the optical disk and with lower density neutral gas at the \ion{H}{1} disk periphery.  We hypothesize that the clusters near the edge of the optical disk are formed in the extension of spiral structure from the inner disk and are a fairly consistent phenomenon and that the clusters formed at the periphery
of the \ion{H}{1} disk are the result of accretion episodes.
\end{abstract}
\keywords{galaxies: star clusters  --  galaxies: structure  --  methods: statistical -- facility:LBT}
]

\section{Introduction}

In the current paradigm of galaxy evolution, material continually accretes onto
galaxies. As such, the outer extremities are particularly interesting environments, but they
are notoriously difficult to study. The surface brightness of the diffuse stellar component drops
well below the background sky level \citep{Pohlen02}, the neutral gas becomes ionized \citep{Maloney93}, and 
observing the diffuse ionized gas requires long exposures on the largest telescopes while
still being limited to within the radial extent of the neutral gas \citep{Christlein08}. 

The GALEX mission \citep{Martin05} highlighted an alternative approach to the study of outer disks
by clearly identifying large populations of apparent young stellar clusters in the outskirts of some nearby
disk galaxies \citep{Thilker05,GildePaz05,Zaritsky07}.
GALEX UV imaging proved particularly useful in this regard because it
provides the stark color contrast required to easily differentiate young, 
blue clusters from the sea of redder background objects.  Of course, any method
that distinguishes between young clusters and background objects can be used to the same effect,
and so, narrowband \Ha\ imaging both predates the GALEX work \citep{Ferguson98} and
still provides new results \citep{Werk10}. While having certain advantages, 
the drawback of the \Ha\ work, and to 
a lesser extent the GALEX studies, is that they focus on the youngest clusters, which
limits the number of such objects available for study and also our ability to measure the long-term
history of this galactic component. Nevertheless, the value of these studies is evidenced by
the increase in corresponding theoretical studies \citep[e.g.][]{Bush08, Roskar08a, Roskar08b, Kazan09}.

\begin{figure*} 
\epsscale{1.3}
\plotone{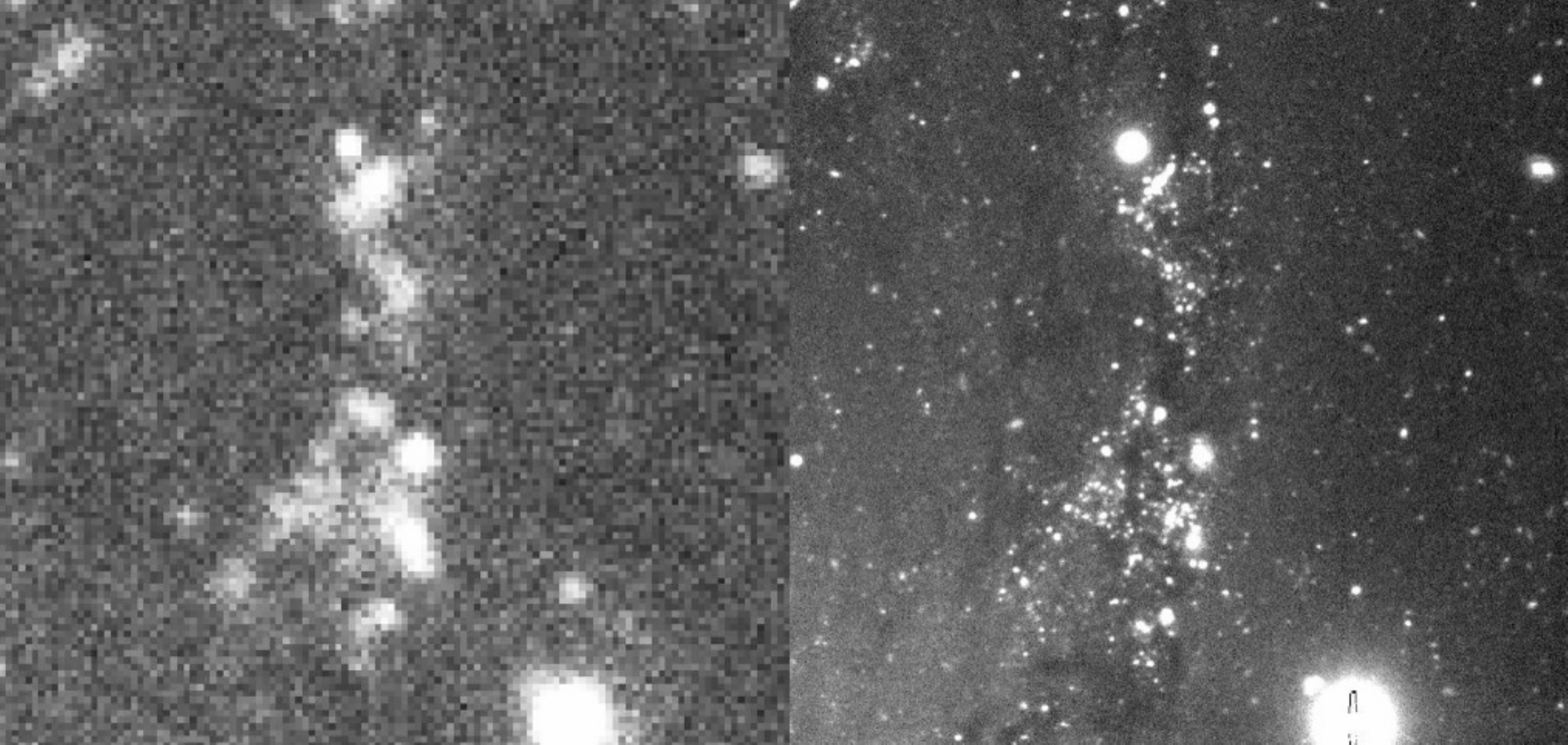}
\caption{Comparison of GALEX (left) and LBT imaging (right) of a 3 arcmin-wide (4.1 kpc) region in the outskirts of
NGC 4736. }
\label{N4736_GALEX_LBT}
\end{figure*}

As useful as GALEX has been for detecting large populations of outer disk clusters, 
this paper focuses instead on broadband optical imaging of outer disks 
for several reasons.  First, the 
spatial resolution of GALEX imaging ($\sim5''$ FWHM) is roughly 
six times poorer than what we normally achieve using ground-based optical telescopes.  
Therefore, the candidate clusters identified by GALEX
are typically blends of multiple clusters \citep{GildePaz05}, as can also be seen
in the comparison we provide in Figure \ref{N4736_GALEX_LBT}.
Until we establish what the typical mass of our detected objects is, we will refer to these objects as stellar knots, which is a purely observational definition of an individual object identified in the available images, to avoid prejudging their true physical nature. Of course, even after we have an estimate of the typical mass one must remain aware that any particular detection may be quite different from the mean.
Second, our knot mass limit is an order of magnitude lower than that of the GALEX samples, leading to a corresponding increase in the number of identifications.  Third, optical colors can, in principle, differentiate stellar populations that are several times older
than those differentiated with the UV colors, thereby extending the 
baseline over which the phenomenon can be studied, and again increasing the sample size.
If we can overcome the difficulty in distinguishing
between stellar knots and background sources in optical images, 
high-quality ground-based data will provide a 
much greater number of knots, and potentially a larger sample of galaxies with such data, 
for any statistical analysis of outer disk properties.

The key to success lies in identifying some property, in addition to color, that can be used to disentangle the stellar clusters from background galaxies. \cite{HF09} used the self-clustering of the outer disk knots to that effect and presented results for NGC 3184.  Here we apply both the ``classical" color approach and our
self-clustering method to characterize these outer disk populations. With results drawn from these
two approaches, we address questions regarding the ubiquity of outer disk knots, measure the knot
formation rate, probe the radial extent of the knot distribution, and compare the distribution of knots to the \ion{H}{1} distribution. 

We present the results of a statistical study of nearby ($<15$ Mpc) outer disks, using 
the 2$\times$8.4m Large Binocular Telescope \citep[LBT;][]{Hill06} 
and wide-field, prime-focus Large Binocular Cameras \citep[LBC;][]{Ragazzoni06,Giallongo08}.  
We describe our data reduction, develop analysis tools, and apply these to 
deep $U$-band and $V$-band imaging data 
of six galaxies.
We demonstrate how the two separate statistical methods enable us to trace outer 
disk cluster-like objects to large radii. 
In \S2 we describe our observations, data reductions and source detections.  In \S3 we present 
color-magnitude diagrams (CMDs) of candidate outer disk sources, quantify the range of  properties consistent 
with the observed knots, and characterize the knot populations around galaxies.
In \S4 we describe the application of our restricted three-point correlation analysis of 
LBT knots, a similar analysis of GALEX knots, and a 
cross-correlation analysis of LBT knots and the underlying \ion{H}{1} distribution.  
First, we present the different analyses, and then discuss them jointly 
on a galaxy-by-galaxy basis in \S5.  
We summarize our results and conclusions in \S6. 

\section{Observations to Final Source Catalogs}

We present new observations of six galaxies (IC 4182, NGC 3351, NGC 4736, NGC 4826, NGC 5474, NGC 6503) 
with the LBC-Blue imaging camera on the LBT on the dates listed in Table~\ref{tab:gals}.  These systems 
comprise a representative sample of nearby disk galaxies by spanning a range of physical disk sizes. All are nearby ($D \sim15$ Mpc) and only two, NGC  4826 and NGC 6503, are 
seen at high inclination ($> 60^\circ$). For the four low inclination systems, their small distances and low inclinations allow 
us to measure the clustering of sources in the disk as 
a function of galactic radius.  We construct the final galaxy mosaics from many individual 
164-second dithered exposures 
through the $U$ and $V$ Bessel filters (hereafter $U$ and $V$), 
except for NGC 5474, for which we combine 33-second dithered exposures 
in $U$ to avoid saturating a bright star in the field.  
The typical seeing during our observations was $0\farcs8$ and never exceeded $\sim1\farcs3$.  
IC 4182, NGC 4736, and NGC 5474 were observed under photometric conditions.  We obtained
single photometric exposures 
for NGC 3351, NGC 4826, and NGC 6503 on 2/8/08 ($U$ \& $V$), 4/20/07 ($U$ \& $V$), and 
4/22/07 ($U$) \& 4/7/08 ($V$), respectively, for photometric calibration.  We observed six 
photometric Landolt standard star fields \citep{Landolt92} for flux calibration and all magnitudes 
are on the Vega system.

We follow the image processing steps described by \cite{HF09}.  
We use a  set of Interactive Data Language (IDL\footnote{Developed by Research Systems, Inc.\ and 
owned by ITT; http://www.ittvis.com/ProductServices/IDL.aspx}) scripts we created to 
construct `master bias' frames that correct for global changes and the two dimensional structure 
in the bias level across the CCDs.  We correct for sensitivity variations across the CCDs by 
combining dithered, twilight-sky flat-field images that are normalized by the four-chip median 
value of each bias-corrected flat field exposure.  We then median-combine the normalized 
flats to create the master flat frames and, finally, divide the bias-corrected science frames 
by the master flats to complete the processing of the individual science exposures.

The target galaxies are well centered on the mosaic, so
we sample the background of each exposure in twenty regions around the edges of the 
detector array. From these twenty regions we are able to obtain sufficient samples from which to judge the uniformity of the background and obtain and estimate of the uncertainty in our background estimate.
The background is estimated using the 
IDL routine \textsc{mmm}\footnote{Part of the Goddard IDL library, maintained by 
W.\ Landsman; http://idlastro.gsfc.nasa.gov/}, which uses an iterative process to determine the 
mode of the sky values after rejecting outlying pixels (caused by stars, saturation, 
cosmic rays, hot pixels and bad columns).  This step is done for each four-chip exposure 
and the minimum estimated sky from the various regions is subtracted from each corresponding chip.  
We then create a weightmap for every four-chip exposure to mask bad columns 
(detected by eye in the master flats), hot pixels and cosmic rays (detected using the 
IDL routine \textsc{reject\_cr}, which finds features sharper than the PSF) when creating the mosaic images.

To combine the individual frames into final exposures we first use SExtractor \citep{Bertin96} 
to create a source catalog for every exposure so that 
SCAMP\footnote{Version 1.4.0; http://terapix.iap.fr/soft/scamp} \citep{Bertin06} can 
correct for optical distortions in LBC-Blue by solving the 
astrometry of the dithered exposures and creating smooth distortion maps for use by 
SWarp\footnote{Version 2.17.1; http://terapix.iap.fr/soft/swarp}.  In cases where the 
SDSS fifth data release \citep[DR5;][]{DR5} does not cover the science field, we use the USNO-B 
catalog \citep{Monet03} as the reference.  We use SWarp 
to mean-combine the spatially-aligned frames after accounting for the bad pixel maps.  Our final 
mosaic images are rebinned to a $\sim0\farcs22$ pixel$^{-1}$ spatial scale and have their 
photometry normalized to a fixed counts arcsec$^{-2}$ second$^{-1}$ standard.  
The mosaics are flat to $0.5 - 1\%$ and 
have total integration times of $\sim25$ minutes in the deepest areas (see Table~\ref{tab:gals}).

We build aperture photometry catalogs using SExtractor, 
with source detection and aperture placement based on the deeper $V$-band mosaic.  
We select sources by identifying groups of five or more pixels each with 
flux $>1\sigma$ above the background (so detections are in total
$>3\sigma$).
Because we detect the integrated light 
from members of stellar groups, the `knots', 
any photometric algorithm that requires a uniform 
object shape for extraction is not optimal for this work.
See \cite{HF09} for all of the chosen SExtractor parameters.  
We detect nearly all 
visually discernable objects beyond the optical radius $R_{25}$, though 
our catalog becomes noticeably incomplete inside $\sim0.8R_{25}$.  Any algorithm will 
have difficulty detecting sources over the bright, extended emission of the inner disk. Issues regarding the completeness relative to our visual detections are negligible because 
the faint objects that are of interest in such a discussion are later rejected on the grounds of the low precision of their color measurement.

We use processed photometric exposures of Landolt standard star fields, taken on the same night as 
our individual photometric exposures of the galaxies, for calibration.  We account 
for an airmass and a color term 
when flux-calibrating on 
the standard Vega system.  Finally, we bootstrap the photometry of the photometric exposures to 
the deep mosaics using $\sim10$ stars common to both.

Colors and magnitudes are measured using circular apertures.  
Colors quoted throughout are from apertures with a diameter fixed to four pixels
($0\farcs9$, just larger than the 
typical $0\farcs8$ FWHM of detected sources), 
while $V$ magnitudes are apertures with a fixed 10 pixel diameter
($2\farcs24$) that are then aperture corrected using stellar curves of growth (see 
Table~\ref{tab:gals} for those corrections, $V_{acorr}$).  Aperture corrections were calculated from $\sim8$ isolated, 
unsaturated stars measured in 15 apertures 
spanning $2-50$ pixels in diameter (or $0\farcs4 - 11\farcs2$).   
We set SExtractor to mask and correct for contaminants.
We check our photometry by comparing the 
$U$ and $V$ apparent magnitudes of ten well-isolated objects across 
the fields with those provided by 
SDSS-DR5 when available, otherwise USNO-B, 
converted from $u$, $g$, and $r$ to either $U$ or $V$ using the 
transformations of \cite{Jester05}.  Aside from the systematic offset 
in our $V$ magnitudes from NGC 4826 (see below), our results are consistent 
with the transformed SDSS photometry to within the transformed 
photometric errors. 

The night of April 24, 2007 had poorer seeing than usual 
($\sim1\farcs3$) during our $V$ exposures of NGC 4826.  As a result, the $U-V$ colors of NGC 4826 
sources, as estimated by eye from the CMDs presented in the next section, appear to be artificially blue by $\sim0.5$ mag relative to the source distributions of the other galaxies.
This shift is as expected given the large aperture correction in $V$ 
(Table~\ref{tab:gals}) and the fact that the colors are aperture-matched (and hence not
aperture corrected).
A similar color offset, though of lower magnitude, is also just noticeable in the 
color-magnitude diagrams of NGC 5474 and NGC 6503 and 
these galaxies have the next highest aperture corrections.
The solution to this problem is to PSF-match all the frames, but that would result
in a significant degradation of the data. We choose not to apply the PSF matching because
these color offsets do not impact our results.
None of the subsequent discussion is predicated on precisely measured colors and
NGC 4826 (the galaxy showing the 
strongest effect), as well as NGC 6503, are eventually
dropped from the analysis because of their relatively high inclinations ($i = 61^\circ$ and $74^\circ$, respectively).

To complete our final source catalogs, we reject sources 
whose internal $U-V$ color error is $> 0.5$ mag (magnitude errors are provided 
by SExtractor and propagated in the standard manner) and
interactively mask regions around bright stars 
(those showing diffraction spikes and halos of scattered light)
because the SExtractor catalogs have 
artificially low counts in those regions, which creates artificial structure in our correlation maps.  
In summary, our catalog is surface brightness limited due to the criterion on flux per pixel, magnitude limited due to the criterion on number of pixels above the particular surface brightness, and color limited due to the requirement of a modest uncertainty in the color measurement. The latter two in particular drive some of the structure seen in the color-magnitude diagrams discussed in \S3. Our catalog is also subject to the confusion limit, although evidently at a finer resolution limit than GALEX (see Figure \ref{N4736_GALEX_LBT}).
We provide 
the number of sources in our final catalogs between $1.0 - 1.5 R_{25}$, 
$1.5 - 2.0 R_{25}$, and $2.0 - 2.5 R_{25}$ and the area in each annulus after masking
(Table~\ref{tab:CMD}). The chosen outer limit in this Table, $2.5 R_{25}$, is somewhat 
arbitrary, loosely based on 
previous outer disk studies, for example \cite{Zaritsky07}.

\section{Outer Disk Cluster Populations}

We will eventually appeal to our clustering analysis to tease out the most information regarding the distribution of the knots in the outskirts of these galaxies. However, it is worthwhile first examining color-magnitude diagrams to gain some intuition regarding the nature of the knots. The aim of this section is therefore not a detailed description of the knots, because we are dominated by uncertainties in accounting for the dominant background population, but rather a broad appraisal of the knots. Estimated masses and ages are highly uncertain.

\begin{figure}
\epsscale{0.7}
\plotone{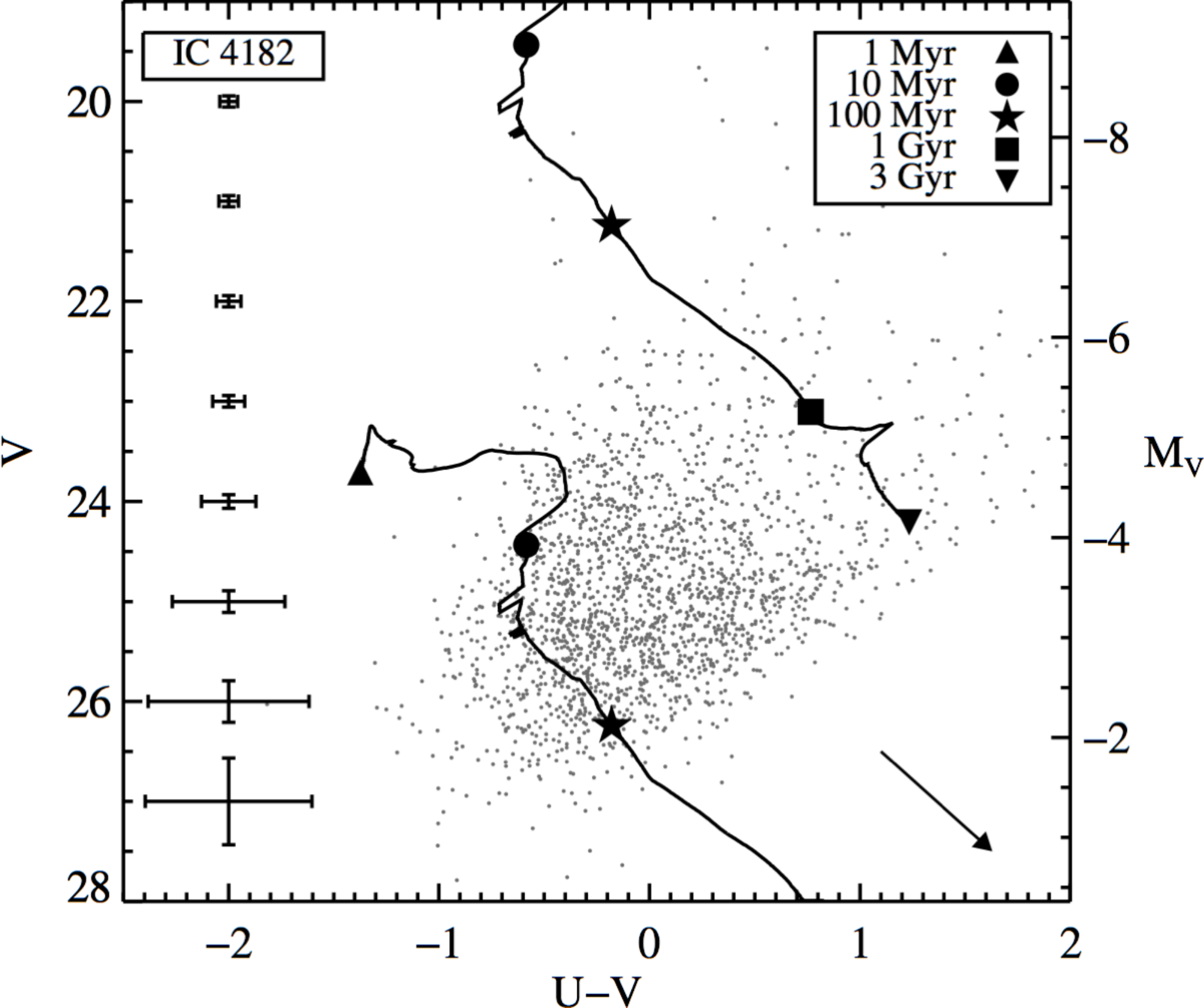}
\caption{CMD of {\it{all}} sources surrounding IC 4182 (including background sources) between $1.0 - 1.5 R_{25}$ 
(after masking areas around bright stars; see text), with 
median 1$\sigma$ errors as a function of apparent magnitude shown on the left.  
A reddening vector corresponding to 1 magnitude of extinction in $V$, calculated 
using results from \cite{Rieke85}, is shown at lower right.  
The work of \cite{Alberts11} suggests low extinction in outer disks 
($E(B-V) < 0.3$).  The tracks represent solar-metallicity $10^3$ (lower) and $10^6 M_\odot$ (upper) star clusters as a function of age. These tracks do not account for stochastic sampling of the stellar mass function. See \S3 for more details.}
\label{IC4182_colmag}
\end{figure}

In Figure \ref{IC4182_colmag},
we present the color-magnitude diagram, CMD, of sources 
between $1.0 - 1.5 R_{25}$ in the field of IC 4182 and tracks
from default Starburst99~\citep{Leitherer99,Vazquez05} that
show the locations of model clusters of fixed mass and
solar metallicity for ages between 1 Myr to 3 Gyr.
We have also examined models with 1/5th and 1/50th solar metalliciites.  The 1/5th solar
model is indistinguishable at the level of precision relevant for the qualitative conclusions we reach ($< 0.5$ mag), and the 
1/50 solar model differs significantly, $> 1$ mag, but only for ages $<$ 10 Myr, at colors where we find few knots in most of our galaxies. Given current observational constrains \citep{Werk11}, we do not expect outer-disk clusters to have extremely sub-solar metallicities.
The upper and lower tracks represent 
$10^4 M_{\sun}$ and $10^2 M_{\sun}$ clusters, respectively, scaled 
from a simulated $10^6 M_{\sun}$ cluster that adequately samples the upper mass range of
a Kroupa IMF.
Because we ignore the stochastic sampling of the IMF at low cluster masses
\citep{Cervino04,Fagiolini07}, a simple comparison between data and the scaled model tracks can lead 
to an underestimate of cluster mass and an overestimate of the age.  Therefore, these tracks 
are only meant to provide a general impression of the cluster masses and ages 
consistent with the range of sources in our sample.

More instructive are the 
background-subtracted Hess diagrams (Figures 3 --- 8) that we
create in the 
same manner as in \cite{HF09}, to statistically constrain the color and magnitude range 
of sources most likely to be clusters associated with each galaxy.  
A Hess diagram is a plot of the 
number of sources within color-magnitude bins 
(we use square bins of 0.2 mag here). For these particular diagrams, 
we create both a `background' Hess diagram from a region far outside the galaxy 
and a `outer disk+background' Hess diagram from the region of interest (say $1.0-1.5 R_{25}$), 
scale the counts in the `background' Hess diagram by the relative areas of the two regions, and subtract 
it,  bin-by-bin from the `outer disk+background' Hess diagram.  The three panels in each Figure show the residual source density between $1.0 - 1.5\ R_{25}$ (left; dark regions are positive counts), 
$1.5 - 2.0\ R_{25}$ (middle), and $2.0 - 2.5\ R_{25}$ (right).  
Solid and dotted black 
contours outline signal above and below the background, respectively, at $>90\%$ confidence level 
(CL), calculated using the low-count, Poisson single-sided upper and lower limits 
from \cite{Gehrels86}.
Because the solid 
contours surround pixels whose individual value is above the background 
at the $90\%$ CL, groups of such pixels 
are detections at a much greater CL.
Overplotted is a $10^3 M_{\sun}$ Starburst99 
model cluster, scaled from a $10^6 M_{\sun}$
cluster, as shown in Figure~\ref{IC4182_colmag}.  See Table~\ref{tab:CMD} for source counts and 
other details of the Hess diagrams, including our estimates of the 
effective surface brightness of the outer disk cluster components.

All six galaxies show excesses in their background subtracted Hess diagrams between 
$1.0 - 1.5\ R_{25}$, roughly tracing the $10^3 M_{\sun}$ cluster track.  The tilted contours in the lower portion of all the panels reflects the color selection (bluer knots are visible to fainter $V$ magnitudes). All galaxies except 
NGC 5474 show suggestive excess between $1.5 - 2.0\ R_{25}$, though the noise is 
noticeably higher.  
We are skeptical, although not dismissive, of apparent excess between $2.0 - 2.5\ R_{25}$ because those 
diagrams are so strongly peppered with oversubtraction (nevertheless, the morphology of signal beyond $R_{25}$ does follow that seen within $R_{25}$). The principal source of uncertainty here are the variations in 
the background population. The effect of this can be seen in the appearance and disappearance of 
regions of oversubtraction, and in the change in location of such oversubtraction, within the various Hess diagrams. In contrast,
the excesses seen in all six galaxies straddle the model track, lending credence to our association of
such a signal with a physical, associated population of sources. 

In Figure~\ref{AVG_Hess} we present the average Hess 
diagram for our six galaxies, with populations combined at constant M$_V$, with and without 
NGC 4736. We single out NGC 4736 because it is the nearest and largest galaxy (in angular extent)
in our sample, resulting in a disproportionate number of detected sources and a less well-determined
background level.
These average Hess diagrams, either with or without NGC 4736, 
show convincing excess out to at least $2.0\ R_{25}$ in that there is more area enclosed within the solid contours than within the dashed contours (alternatively, based on Table \ref{tab:CMD}, each galaxy has an average excess of over 70 knots between 1.5 and 2 $R_{25}$, even if we exclude NGC 4736). We conclude that there is indeed a population of knots extending well beyond $R_{25}$ in all of these galaxies and that the knots are consistent with a population of stellar clusters. We will now refer to these objects as stellar clusters. The exact distribution of masses and ages is difficult to disentangle from these plots given the uncertainties in the luminosities and colors of low mass clusters arising from the stochastic sampling of the stellar mass function, the uncertainties in the modeling, and the large uncertainties arising from the background subtraction.

\subsection{An Estimate of the Cluster Formation Rate}

To gain
some intuition on the implied cluster formation rate from these diagrams and to further test our assertion that these are clusters, 
we simulate a population of $10^3 M_{\sun}$ clusters forming at a specified constant rate over the previous several Gyr 
and plot the resulting Hess diagrams, accounting for the photometric uncertainties as estimated from the data for NGC 4736 (but not the uncertainties in \hfil

\begin{figure*} 
\label{fig:IC4182_Hess}
\center{
\epsscale{1.8}
\plotone{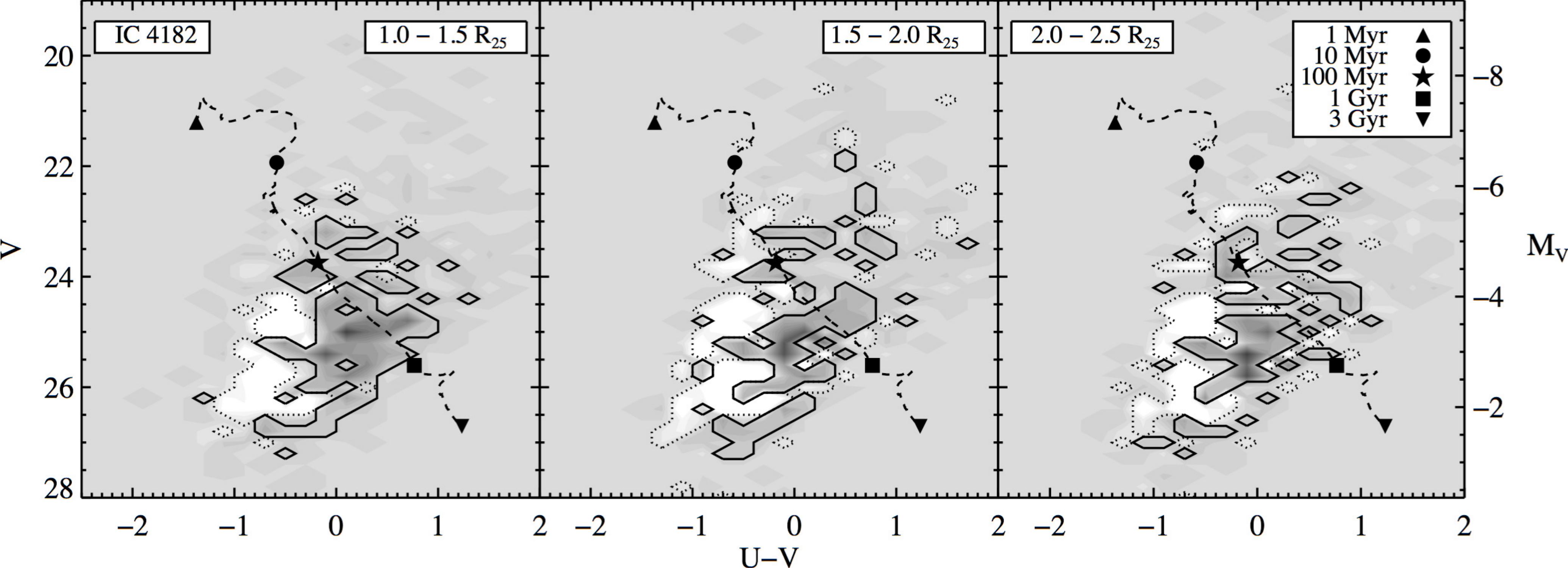}
\caption{Background-subtracted Hess diagrams for IC 4182, made from sources between 
$1.0 - 1.5\ R_{25}$ (left), $1.5 - 2.0\ R_{25}$ (middle), and $2.0 - 2.5\ R_{25}$ (right).  
Dark regions represent positive counts.  Solid and dotted black 
contours outline signal lying above and below the background at a $>90\%$ confidence level 
(CL), respectively; the dotted contours show any oversubtraction.  
Overplotted is a $10^3 M_{\sun}$ Starburst99 
model cluster, scaled down from a $10^6 M_{\sun}$ cluster. Symbols represent the age
of the cluster along the model sequence.}}
\end{figure*}

\begin{figure*} 
\label{N3351Hess}
\center{
\epsscale{1.8}
\plotone{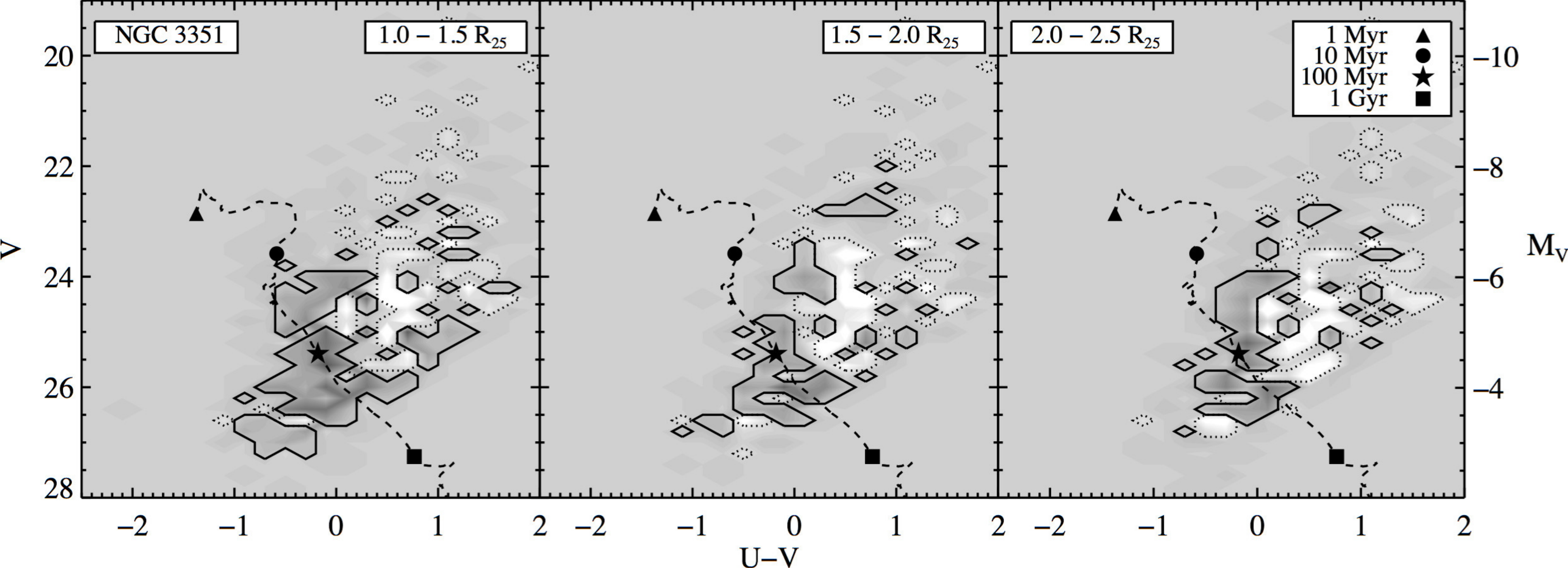}
\caption{Same as Figure 3 but for NGC 3351. }}
\end{figure*}

\begin{figure*} 
\label{M94Hess}
\center{
\epsscale{1.8}
\plotone{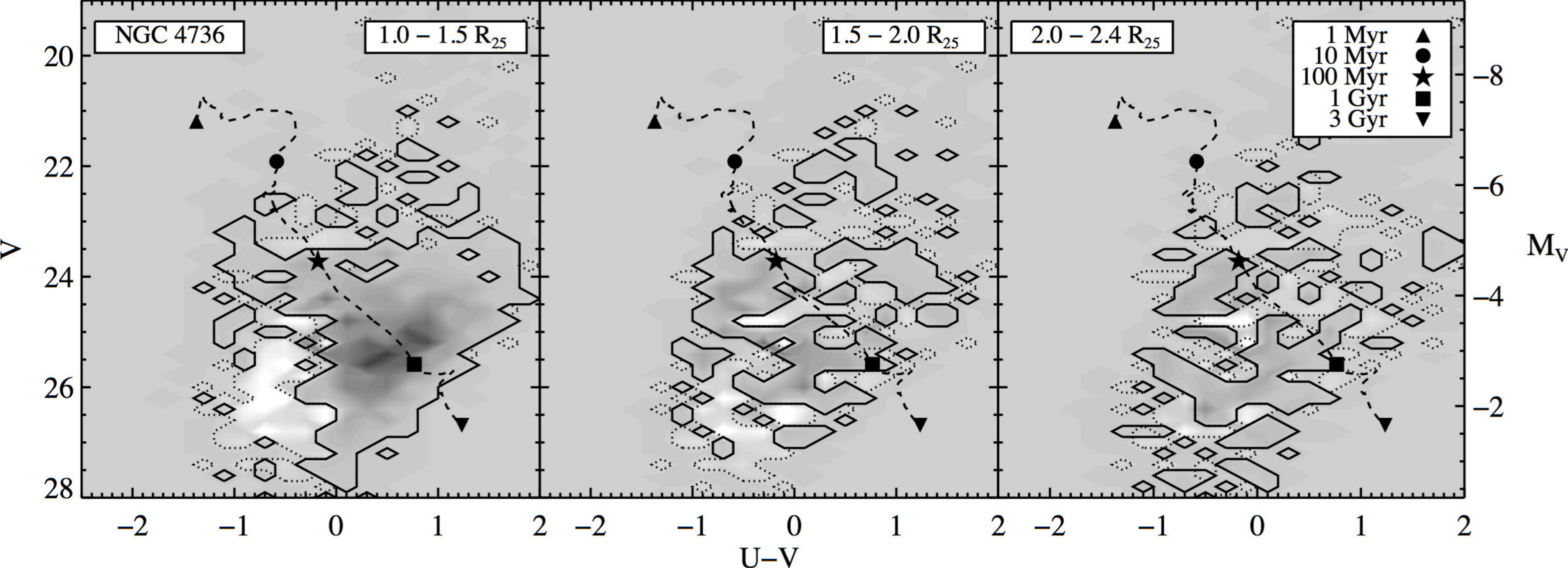}
\caption{Same as Figure 3 but for NGC 4736. }}
\end{figure*}
\clearpage

\begin{figure*} 
\label{N4826Hess}
\center{
\epsscale{1.8}
\plotone{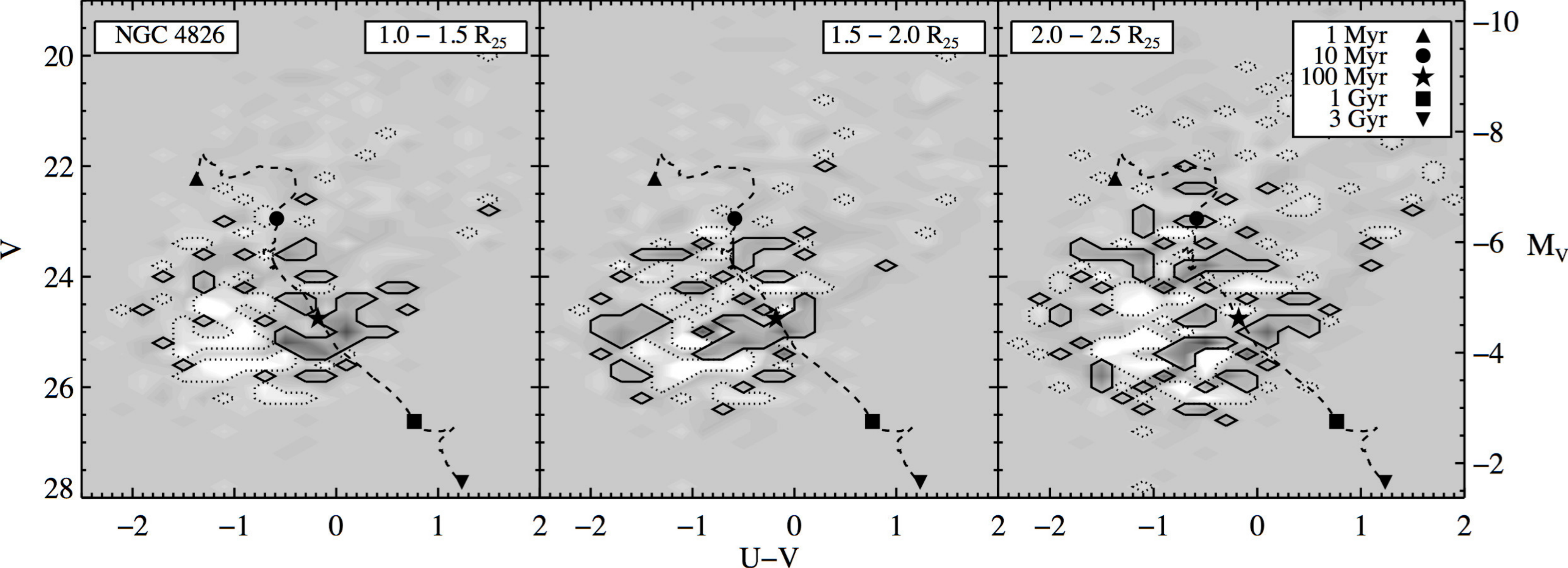}
\caption{Same as Figure 3 but for NGC 4826. }}
\end{figure*}

\begin{figure*} 
\label{N5474Hess}
\center{
\epsscale{1.8}
\plotone{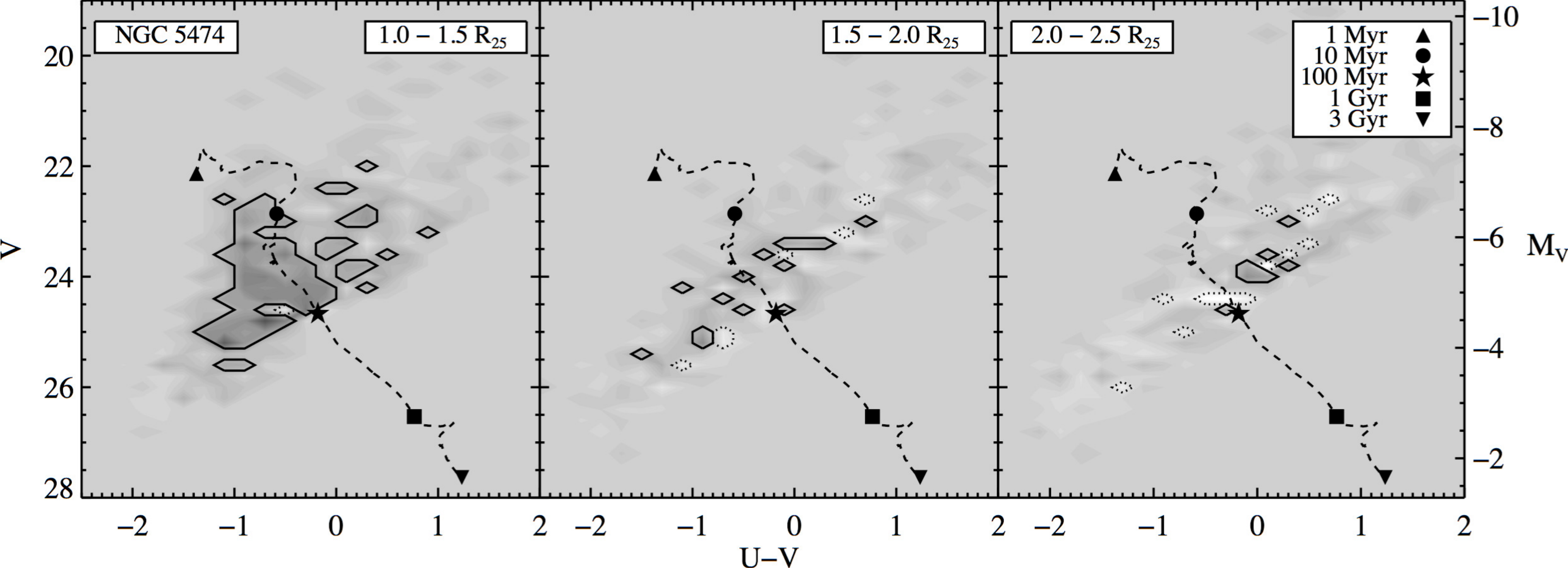}
\caption{Same as Figure 3 but for NGC 5474. }}
\end{figure*}

\begin{figure*} 
\label{N6503Hess}
\center{
\epsscale{1.8}
\plotone{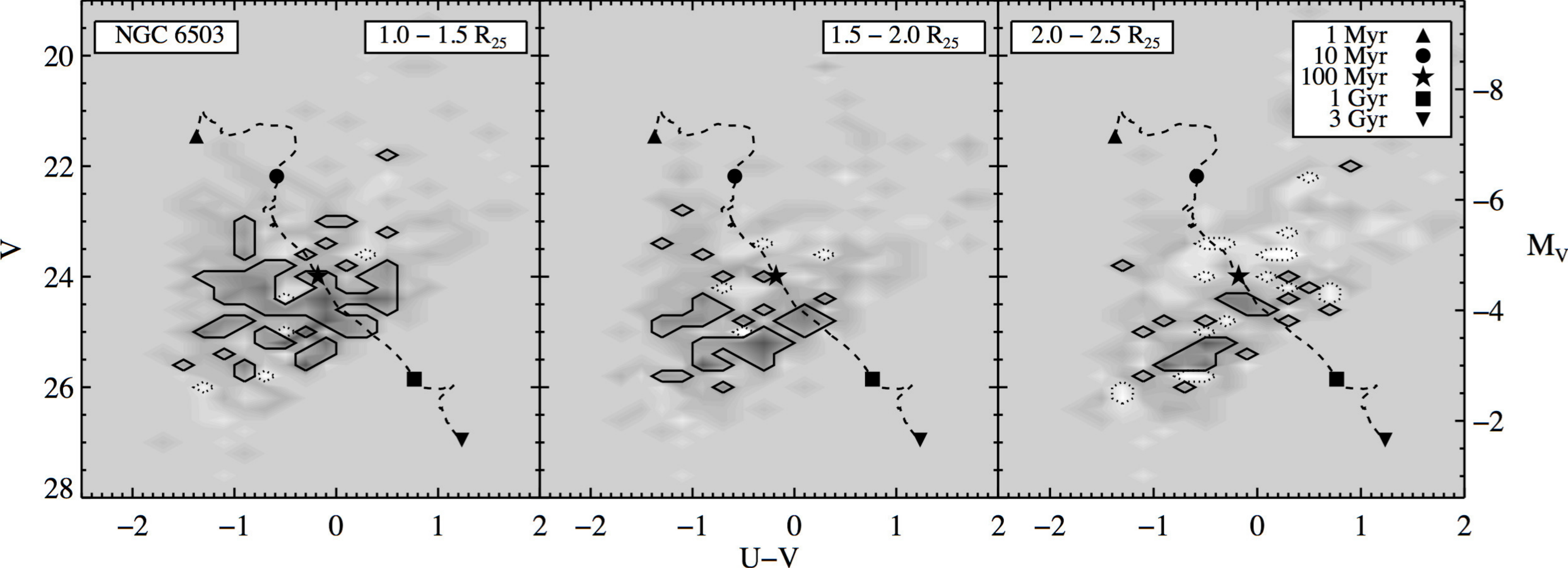}
\caption{Same as Figure 3 but for NGC 6503.}}
\end{figure*}
\clearpage

\begin{figure*} 
\includegraphics[angle=90,width=7.0in]{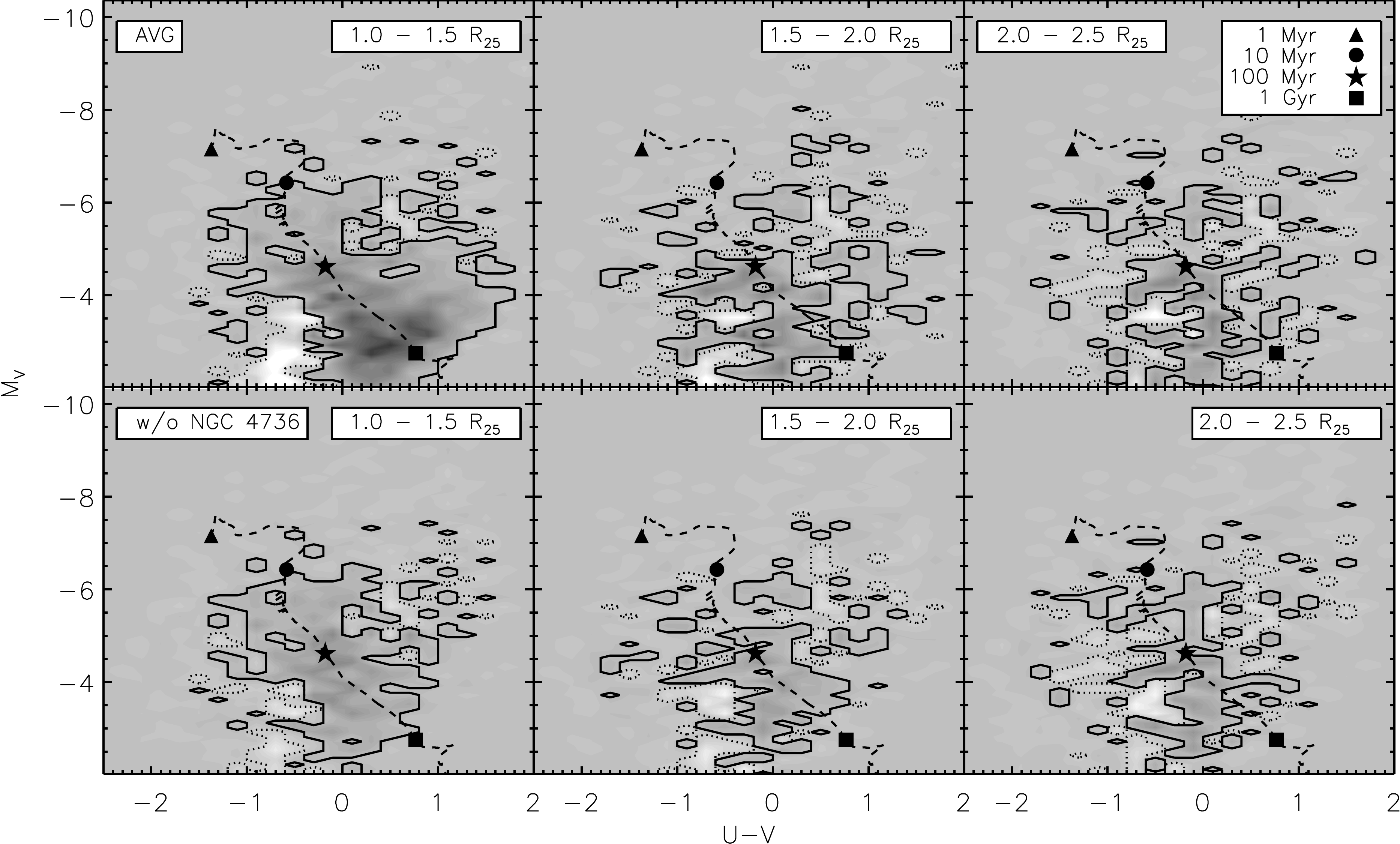}
\caption{The average Hess diagram for all six galaxies, with sources combined as a function of M$_V$, 
including and excluding NGC 4736 (top and bottom panels, respectively).  The three radial ranges 
are the same as in the individual Hess diagrams ($1.0 - 1.5\ R_{25}$, $1.5 - 2.0\ R_{25}$, 
and $2.0 - 2.5\ R_{25}$).  Dark regions are positive counts.  Solid and dotted black 
contours outline signal lying above and below the background at a $>90\%$ confidence level 
(CL), respectively; the dotted contours show any oversubtraction.  
Overplotted is a $10^3 M_{\sun}$ Starburst99 
model cluster, scaled down from a $10^6 M_{\sun}$ cluster.
There is excess signal out to at least $2\ R_{25}$.}
\label{AVG_Hess}
\end{figure*}
\clearpage

\begin{figure*} 
\center{
\includegraphics[angle=90,width=6.0in]{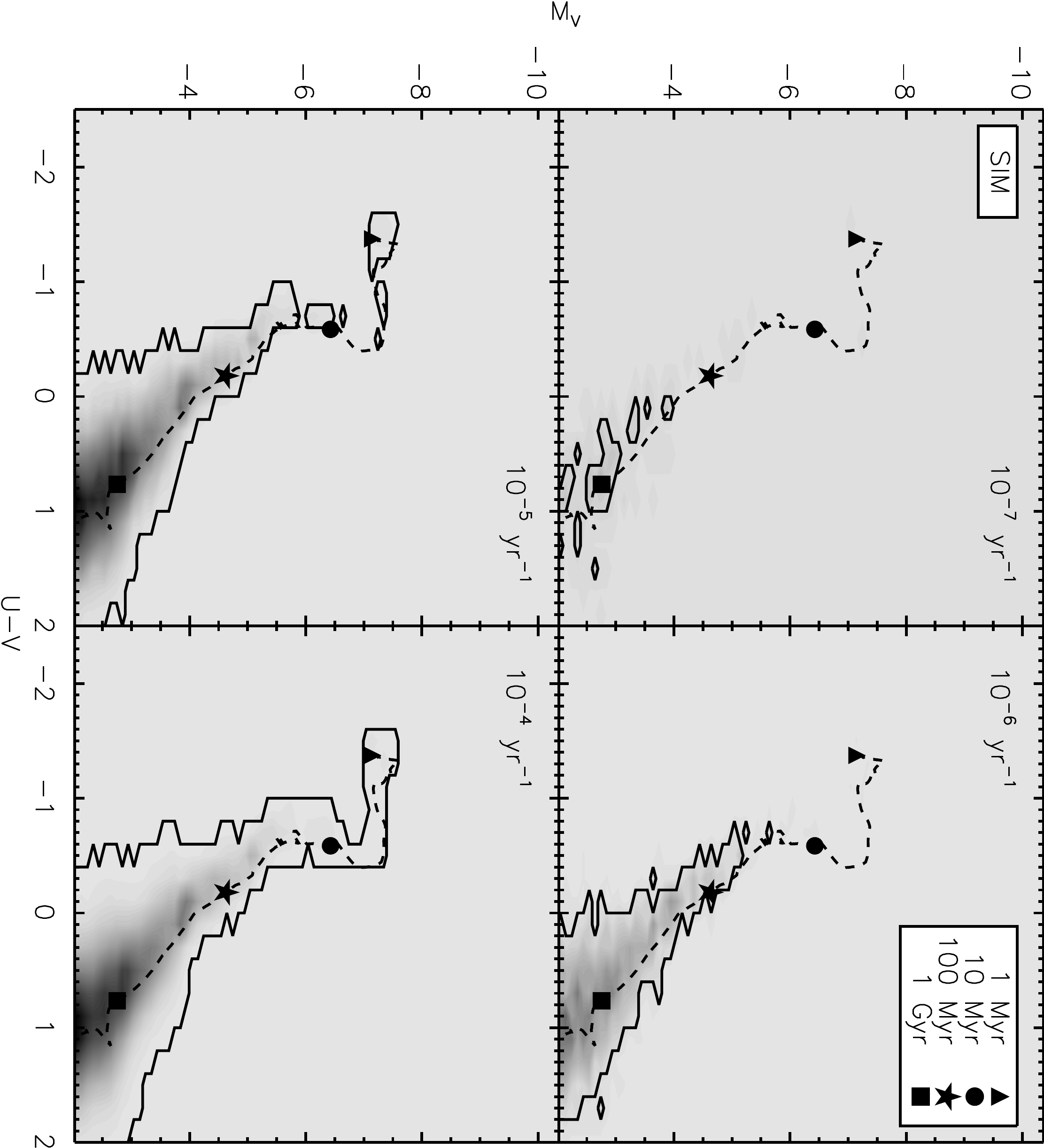}
\caption{Simulated Hess diagrams made from a Starburst99 $10^3 M_{\sun}$ cluster track, with
CMD uncertainties corresponding to those found for NGC 4736, for different cluster formation rates 
(top left: one every $10\sci6$ yr;  top right: one every $10^6$ yr;  bottom left: one every $10^5$ yr;  
bottom right: one every $10^4$).  No background is considered when making the simulated diagrams.  }}
\label{SIM_Hess}
\end{figure*}
\clearpage

\noindent
the background subtraction nor the stochastic sampling of the stellar mass function, in a figure analogous to Figure~\ref{AVG_Hess} 
for four different cluster formation 
rates. We assume no cluster dissolution or disruption, so this comparison will provide a lower limit
on the cluster formation rate. Given the uncertainties in applying these models to the data, arising from the causes outlined above, this calculation is intended only as a plausibility argument for relating the knots to stellar clusters.

Comparing Figures \ref{AVG_Hess} and 10, we conclude that 
the data exclude cluster formation rates significantly higher than one every $\sim10^6$ years for clusters 
of mass $\ge 10^3 M_{\sun}$ (if there is no significant cluster dissolution) 
because we do not detect a population of blue sources at $M_V \lesssim -7$ mag. This conclusion is relatively insensitive to the issue of stochastic sampling because that becomes less of a factor for the more massive clusters. On the other
hand, rates significantly lower than $10^{-6}$ yr$^{-1}$ fail to produce a sufficiently significant population of sources.
This range of rates is consistent with a calculation where we simply take the 
number of outer disk knots in the $1.0 - 1.5\ R_{25}$ Hess diagrams ($\sim400$), the maximum 
age of a $10^3 M_{\sun}$ cluster in our diagrams ($\sim1$ Gyr), and a uniform rate of formation 
over that time, which results in a rate estimate of one cluster every $\sim2.5$ Myr ($4\times 10^{-7}$ yr$^{-1}$). 
Converting this cluster formation rate to a stellar cluster mass rate implies that 
$\sim0.004 M_{\sun}$ pc$^{-2}$ Gyr$^{-1}$ is being tied up in stellar clusters at these radii (assuming $R_{25} = 5$ kpc).

Alternative estimates of the outer disk cluster formation rate exist for comparison.  \cite{Ferguson98}, using 
deep \Ha\ imaging, measured the outer disk star formation rate densities of NGC 628, 
NGC 1058 and NGC 6946 to be between 
$\sim0.01 - 0.05 M_{\sun}$ pc$^{-2}$ Gyr$^{-1}$.
There are at least four potential explanations for our significantly ($10\times$) lower formation rate, 
all of which probably contribute to the difference: 
1) the cluster formation rates are highly stochastic and \cite{Ferguson98} happened to catch these galaxies in an elevated 
phase relative to ours, 2) clusters disassociate, and so many are missing in our sample which samples
older clusters, and 3) the 
\Ha\ technique, which is sensitive to low mass clusters, is measuring a different mass
component that happens to contain a larger fraction of the total mass, and 4) our estimates of the masses are significantly corrupted by our neglect of the stochastic effects and modeling uncertainties.

Regarding the 
first possibility, we know from a comparative study of GALEX knots \citep{Zaritsky07} 
that only a fraction of galaxies ($\sim25\%$) show significant overdensities of bright blue knots.
If all galaxies have an outer disk population, as we seem to find on the basis of
optical imaging presented here and spectroscopy \citep{Christlein08}, then the formation rate 
must be highly variable (with a duty cycle of about $25\%$ for GALEX-detectable knots). The
effects of this stochasticity should be even more dramatic in H$\alpha$.
Regarding the second option, we know that in certain environments where we have 
clusters spanning a range of ages, and can therefore do the study, that  only a small fraction 
of all star clusters survive and we believe we understand the physical mechanism
for this evolution \citep[see][and many related studies]{Spitzer58,gieles}. Depending on the driver for cluster 
dissolution (mass loss versus tidal stresses), the rate of cluster dissolution may be lower 
in the outer disks, but unlikely to be negligible.  \cite{Davidge11} find that clusters in the 
outer disk of M33 dissipate on a time scale of 100 Myr.  
The clusters we detect are $\sim$100 Myr old and older, and so  
likely to be a remnant population. This possibility is given further support by the recent work
of \cite{Alberts11}, who in a set of five galaxies in which they are able to measure an age distribution of 
a set of massive ($M \sim 10^5 - 10^6 M_\odot$), outer disk clusters, find that the age distributions
are all peaked toward early times ($\sim 100$ Myr and often within their innermost age bin of 50 Myr) 
even though they sample to ages of 1 Gyr. 
Finally, regarding the third option, 
\cite{Davidge11} calculate that most of the outer disk 
clusters in M33 form at lower masses ($50-250 M_{\sun}$) 
and so it may be the case that many of the \Ha-detected 
clusters are of similar masses and below our detection threshold. 
It may also be the case that many of our clusters are low mass and boosted by stochastic effects into detectability. As such, any quantitative determination of the mass function of these clusters will require simulations that include detailed treatments of such effects, as well as of dynamical evolution and selection.
Importantly, if we are missing
90\% to 95\% of the outer-disk cluster formation, as suggested by the \cite{Ferguson98} results, 
then we must bear in mind that all of our
subsequent estimates of stellar mass densities at these radii need to be multiplied by a factor of 10 to 20. For now, we ascribe the discrepancy to one of the effects described above rather than to a wholesale missing population from our catalog.

If the outer disks in our sample account for a star formation rates of $4\sci{-4}$  M$_{\sun}$ yr$^{-1}$
(one $10^3 M_{\sun}$ cluster every 2.5 Myr for $1.0 R_{25} \le R \le 1.5 R_{25}$), 
then $\sim4\sci6 M_{\sun}$ of stars have formed in that annulus 
over the lifetime of the galaxy (taken as $10^{10}$ yr).  
As we discussed above, this is a lower limit since it ignores cluster dissolution and potential selection effects. If we multiply this number by 20, then we conclude that this limited region of the outer disk could contain as much as $10^8$ M$_\odot$ of stars, or about 1\% of the stars in a typical large spiral. 

We conclude that adopting rough estimates of the typical mass and age of these clusters results in an outer disk star formation rate that is consistent with other estimates in that it lies below those other estimates. As such, our claim that these are indeed outer disk star clusters does not conflict with those other observations.

\section{Clustering in the Outer Disks}

The analysis of the color-magnitude diagrams are limited by the immense contamination from background objects. As we mentioned before, while at UV wavelengths that contamination is held in check, at optical wavelengths another method is needed to help differentiate outer disk clusters from background objects.
To enhance the contrast between outer disk clusters and the background, we now
utilize spatial correlations among the clusters and between the clusters and other detected,
outer-disk components. First, we describe the self-correlation of LBT-knots. We will discuss the results for individual galaxies in detail, and the sample as a whole, in \S5. Second, we present the application of the same technique to existing GALEX data for these same galaxies. In general, there are fewer GALEX knots per galaxy, so the statistical information is poorer, but those data offer an independent check on our LBT results. Finally, we cross-correlate the position of the LBT-knots with the \ion{H}{1} in an attempt to confirm tenuous evidence for very distant clusters through their association with neutral hydrogen at these large radii.

\subsection{Self-clustering of LBT Knots}

\begin{figure}[h]
\epsscale{0.5}
\plotone{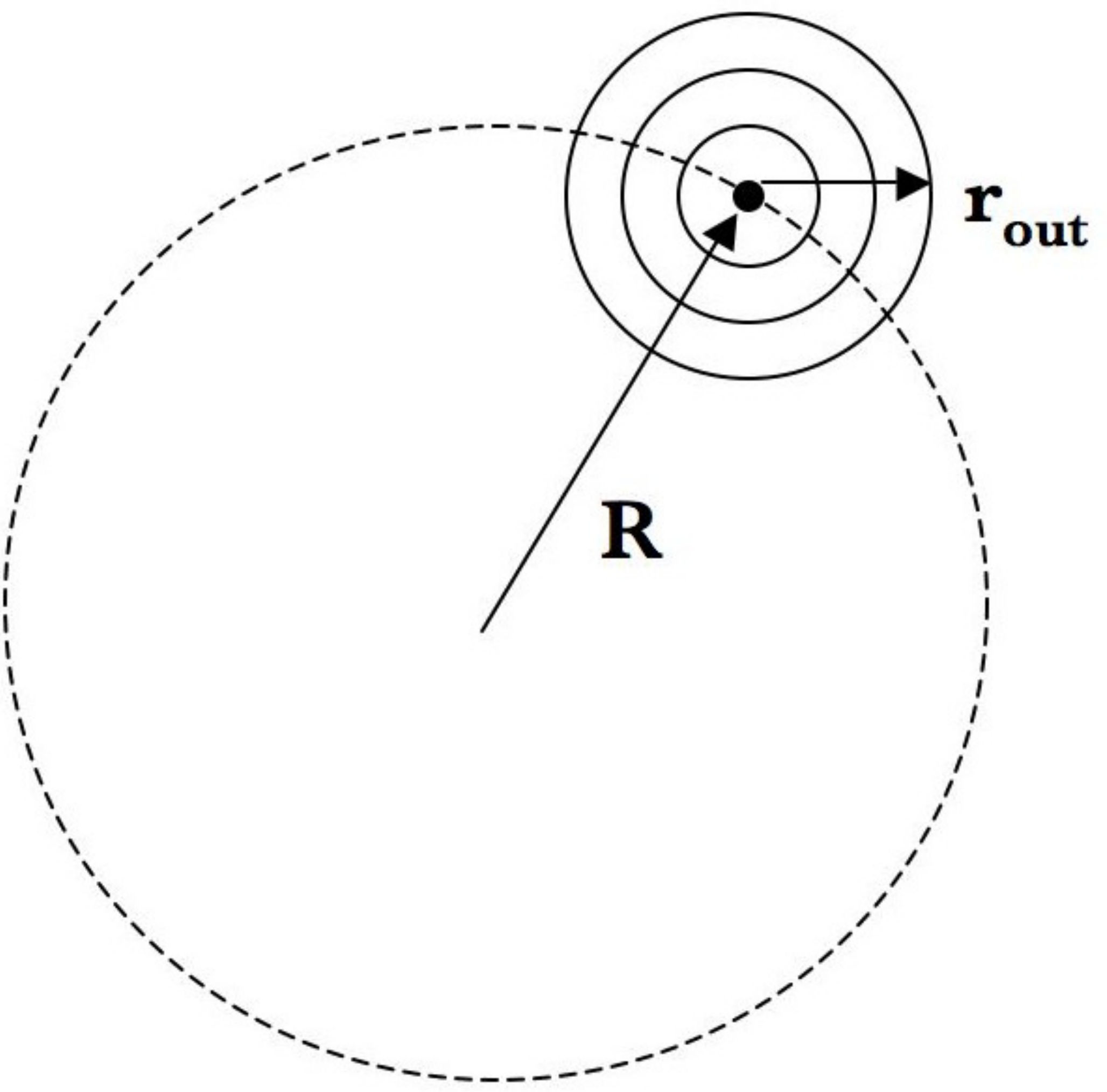}
\caption{The radii defined for creating the restricted three-point correlation maps. The center of the parent
galaxy lies at the origin and the dot represents one outer disk knot.  The original description of this method, along with this figure, are presented by \cite{HF09}.}
\label{radii_expl}
\end{figure}

Following \cite{HF09}, we present restricted three-point correlation maps to trace the self-clustering 
of knots in the outer disks (see Figure~\ref{radii_expl} for a description of the radii used).  
The self-clustering of outer disk clusters provides an enhanced contrast relative to the 
background (which has a different angular correlation function).
Instead of measuring a radial profile of detected sources and subtracting some average background level, which is simply related to the azimuthally-averaged 
two-point correlation function \citep[e.g.][]{Zaritsky07}, here we use the 
self-clustering of knots to highlight regions with clusters and remove 
signal from large-scale background fluctuations.  
This technique only measures the extent of {\it{clustered}} knots; stars or clusters in a diffuse component, even
if originally born in clustered clusters, 
will evade detection. Therefore, we stress that any 
null detection does not indicate a lack of stellar populations at large galactic radii.

\begin{figure} 
\includegraphics[scale=0.7,angle=90]{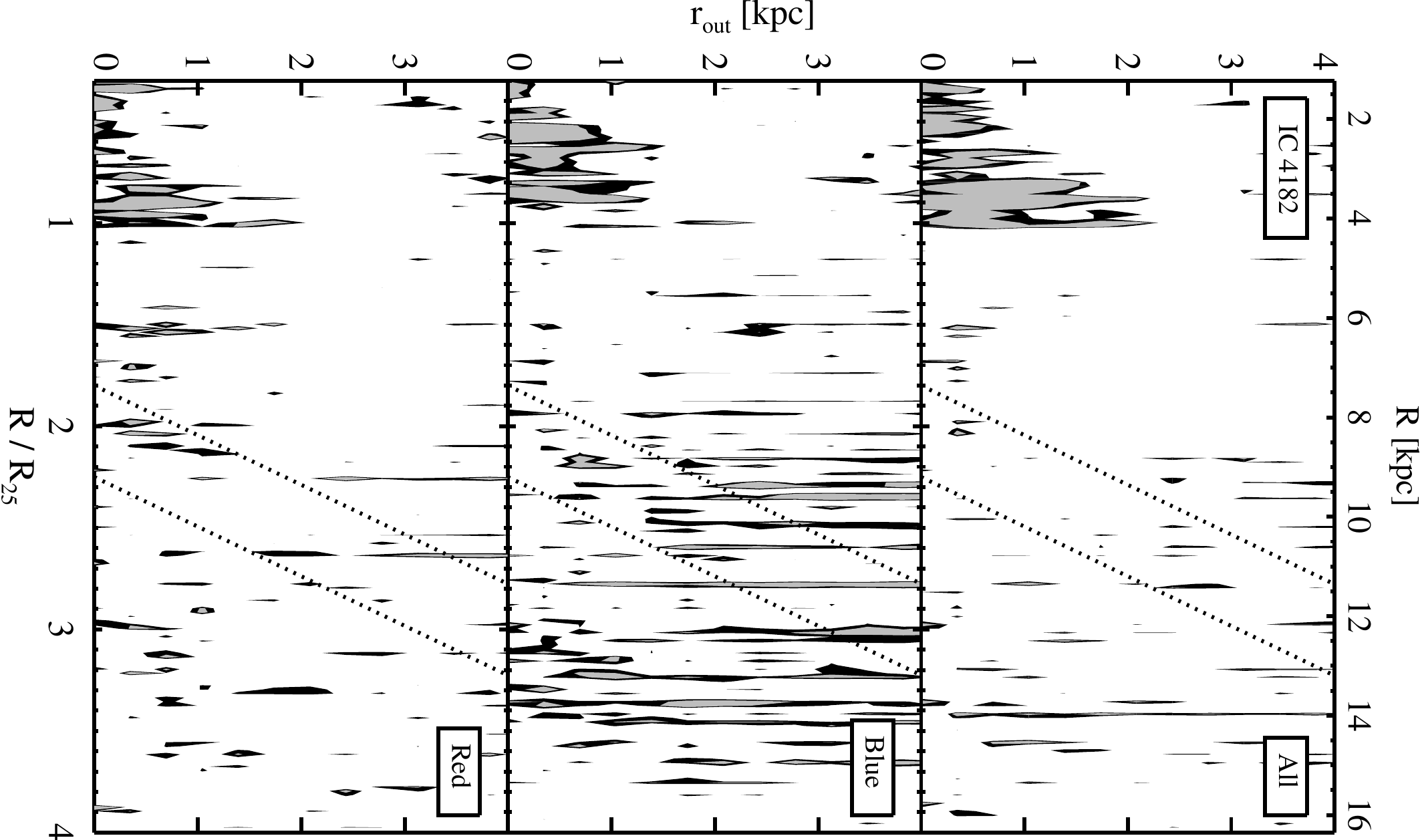}
\caption{Restricted three-point correlation maps from sources in our (masked) final catalog of IC 4182.  
Black and gray 
show areas where signal is detected at the $>95\%$ and $>99\%$ significance level, 
respectively, as a function of both galactic radius $R$ and intercluster radius $r_{out}$ 
(Figure~\ref{radii_expl}).  
The dotted lines show the radial extent of \ion{H}{1} 
for $N$(\ion{H}{1}) $> 2\sci{20} \cm{-2}$ (left, at $r_{out} = 0$ kpc) and for $N$(\ion{H}{1}) above the 
noise level of the integrated \ion{H}{1} map (right, at $r_{out} = 0$ kpc).  The dotted lines distinguish the 
regions where knots could be
co-located with the \ion{H}{1} (left and right of the dotted lines, respectively).}
\label{IC4182_LBT_corr}
\end{figure}

\begin{figure} 
\includegraphics[scale=0.7,angle=90]{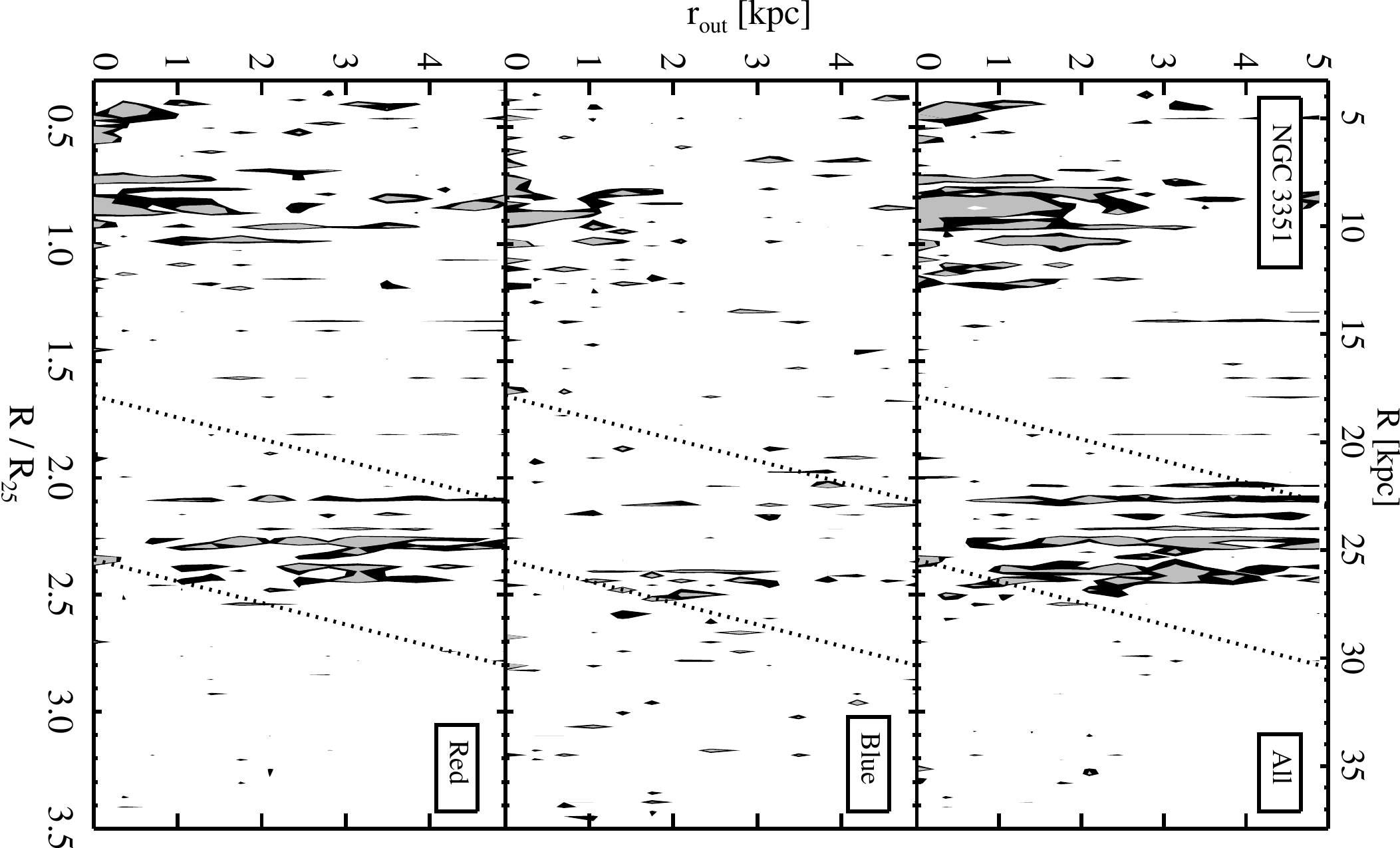}
\caption{Same as Figure~\ref{IC4182_LBT_corr} but for NGC 3351. }
\label{N3351_LBT_corr}
\end{figure}

\begin{figure} 
\includegraphics[scale=0.7,angle=90]{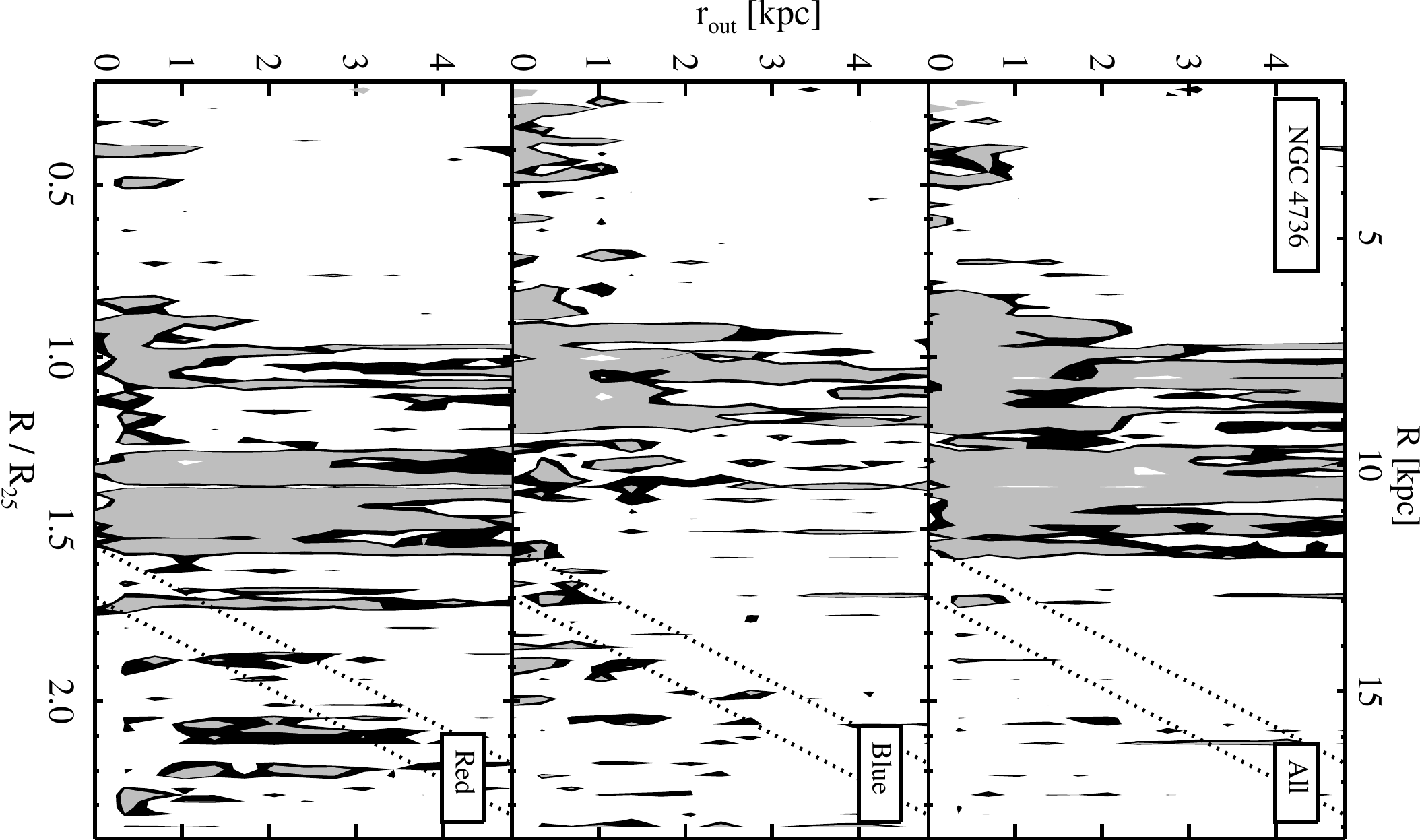}
\caption{Same as Figure~\ref{IC4182_LBT_corr} but for NGC 4736. }
\label{M94_LBT_corr}
\end{figure}

\begin{figure} 
\includegraphics[scale=0.7,angle=90]{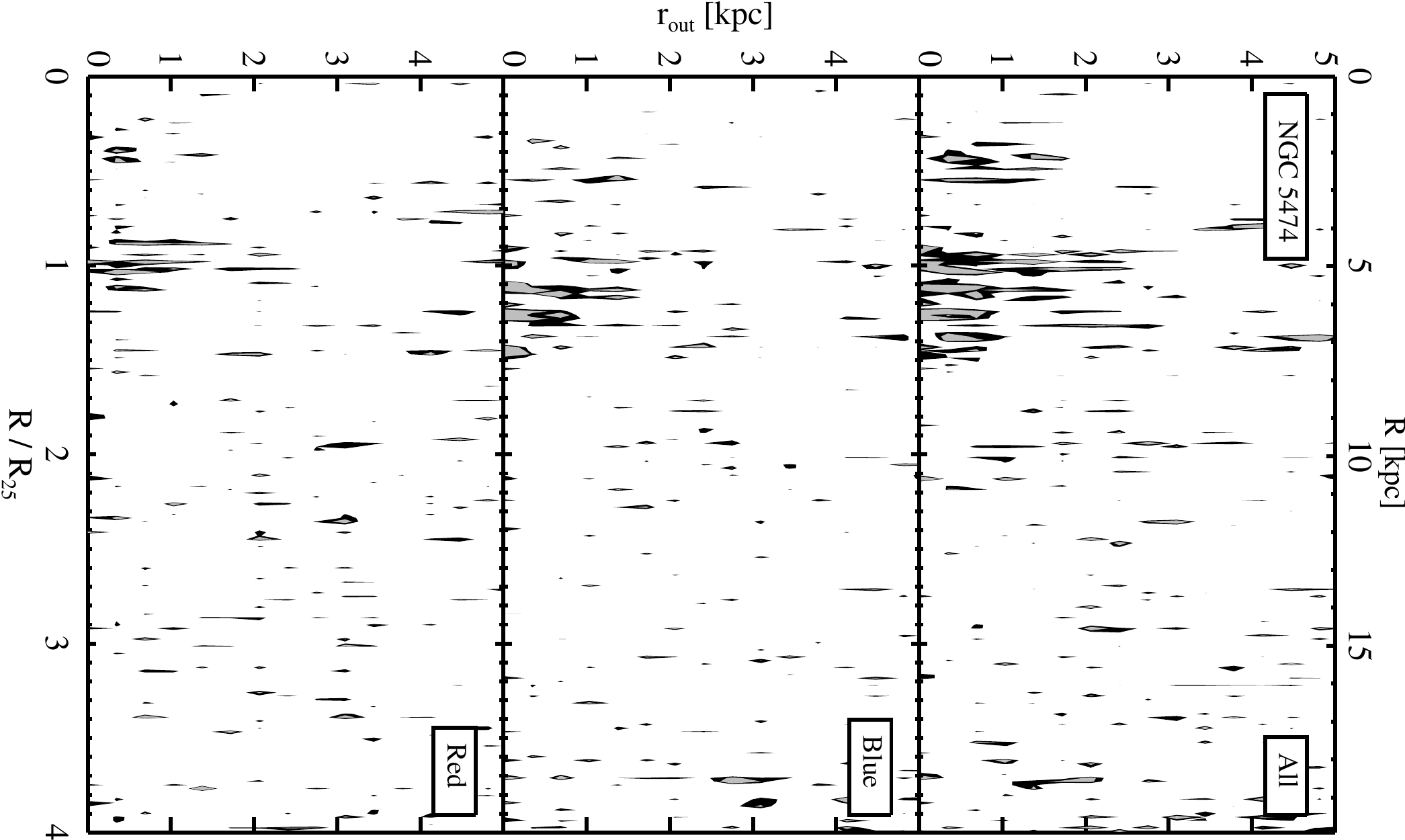}
\caption{Same as Figure~\ref{IC4182_LBT_corr} but for NGC 5474. }
\label{N5474_LBT_corr}
\end{figure}

The restricted three-point correlation maps of our four 
low-inclination galaxies ($i < 60^\circ$) are presented in Figures~\ref{IC4182_LBT_corr} ---  \ref{N5474_LBT_corr}.  
The top panels are constructed using all detections (the `All' sample)
with $-1.7 < U-V < 0.7$ and $19 < V < 27.5$, while the 
lower panels result from splitting the samples into blue and red components on either side of 
$U-V = -0.2$ (middle and bottom panels, respectively, and comprise the `Blue' and `Red' samples). 
Black and gray 
show areas where signal is detected at the $>95\%$ and $>99\%$ significance level, 
respectively, as a function of both galactic radius $R$ and intercluster radius $r_{out}$ 
(Figure~\ref{radii_expl}).  The confidence levels are determined as described by 
\cite{HF09} using a Monte-Carlo approach.
The dotted lines show the radial extent of \ion{H}{1} 
for $N$(\ion{H}{1}) $> 2\sci{20} \cm{-2}$ (left, at $r_{out} = 0$ kpc) and for $N$(\ion{H}{1}) above the 
noise level of the integrated \ion{H}{1} map (right, at $r_{out} = 0$ kpc).  The dotted lines are slanted to 
distinguish the [$R, r_{out}$] regions that can be populated by sources within
the \ion{H}{1} disk.
The \ion{H}{1} data and analysis are described later in the paper.

The interpretation of these Figures is somewhat unusual so we outline the salient features. Positive signal at any location is potentially a marker of clustering. However, positive signal at low $r_{out}$, which is typically seen interior to $R_{25}$ is a sign of small-scale, tight clustering of stellar clusters. Vertical bands, which are seen at a variety of radii, suggest a set of clusters at a particular radius that may not be strongly clustered, such as those in a spiral arm or ring. Because of the nature of the axes, a fixed amount along the vertical axis represents a much larger angle at a smaller $R/R_{25}$ than at a larger one, but we expect clustering to depend on physical separation rather than on angular separation.

Reaching conclusions regarding the radial
extent of correlated clusters from these 
maps is difficult because of the irregular nature of the signal and the statistical
fact that given enough pixels some will randomly host 3$\sigma$ variations.  Rather than 
judge the significance of correlation peaks by eye, we
calculate at each radius whether the occurrence of outliers is statistically beyond what is
expected. For this calculation, we use probabilities calculated using binomial statistics
with the appropriate probability corresponding to that of the unlikely event (for example,
we calculate the likelihood of having as many as $M$ or more pixels out of $N$ 
pixels hosting `true' values if the probability of obtaining a true value is 0.1). 
We rebin our data in the $R$ direction so that each element now corresponds to a $\Delta R$
of the same size as the $\Delta r_{out}$, 
so that each element is independent.

We calculate the probability of finding as many `significant' detections in the correlation maps by chance
as a function of galactocentric radius, over the range of 
$r_{out}$ plotted in the correlation maps ($\sim5$ kpc), and present the results in 
Figures 16 --- 19.
The black and 
gray lines correspond to the black and gray signal in the correlation maps.  The dotted 
lines indicate the same \ion{H}{1} extents as in the correlation maps.  Black or gray 
signal below the dashed line at $-$2.568 indicates concentrations of correlation signal that are 
highly unlikely ($>3\sigma$) to be generated by chance (i.e.\ the correlation signal 
excess is statistically significant).

Before discussing the results of our LBT correlation maps and the corresponding probability 
plots, 
we present a similar correlation analysis of GALEX knots around these same galaxies 
and a cross-correlation analysis of our LBT knots and the underlying \ion{H}{1} distribution.
We then discuss the results of these three analyses 
together, on a galaxy-by-galaxy basis.
We do this because comparison to independent data provides additional credence to our detections.

\subsection{Self-clustering of GALEX Knots}

We now apply our three-point correlation analysis to the distribution of UV-bright knots 
around our galaxies, using data from publicly available GALEX catalogs.  If the outer disk regions that we detect in the LBT data are continuing to form clusters, then we might detect corresponding GALEX knot self-clustering (i.e.\ GALEX knot - GALEX knot clustering). Such a detection would not only
confirm what may be marginal detections in the LBT data, but also provide information on the
timescales of spatially localized cluster formation and cluster dispersal. 

\begin{figure} 
\center{
\includegraphics[scale=0.4,angle=90]{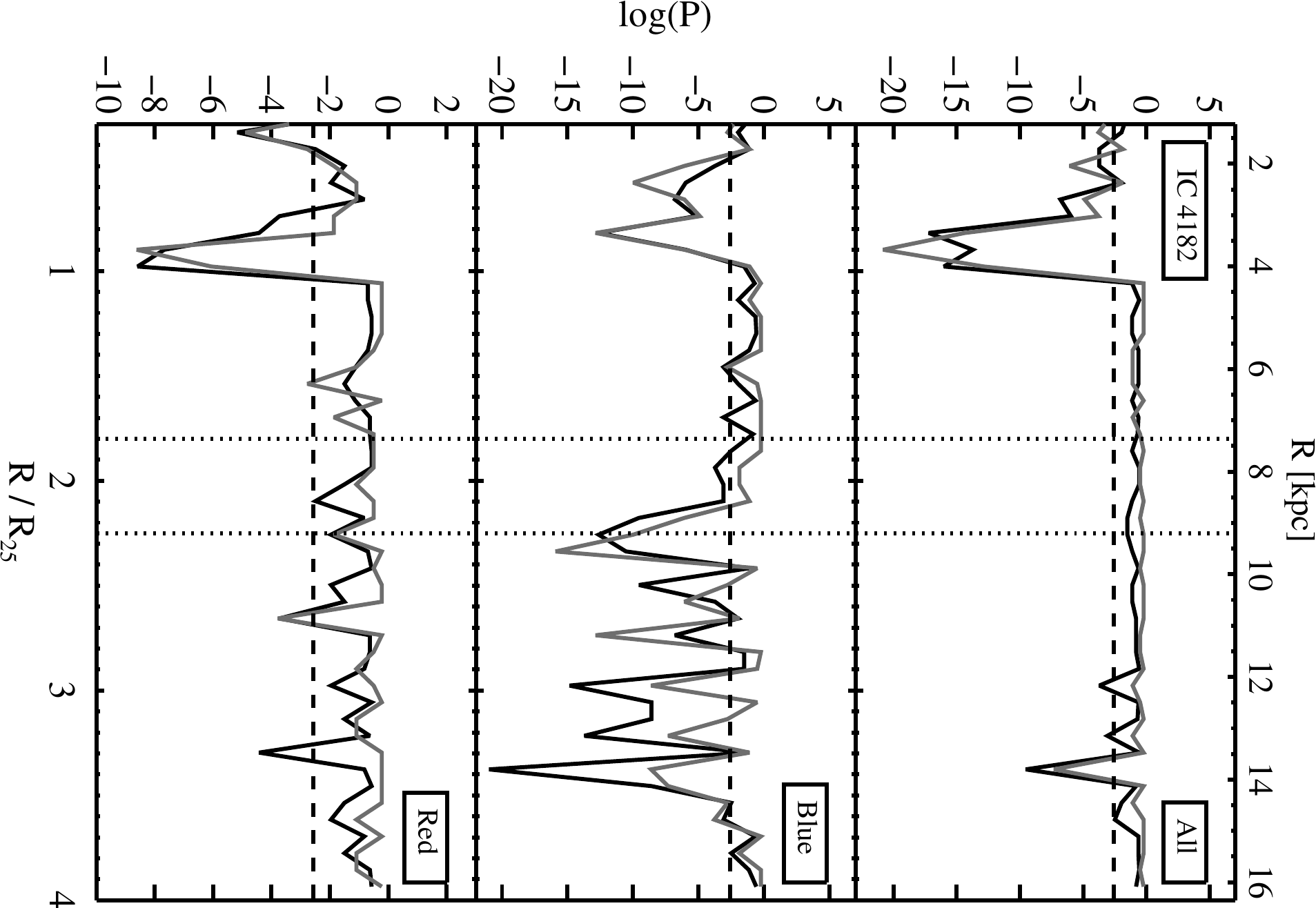}
\caption{Probability, as a function of radius, that signal in the three-point correlation map of 
IC 4182 (Figure~\ref{IC4182_LBT_corr}) is due to random excursions.  The black and 
gray lines correspond to the black and gray signal in the correlation map.  The dotted 
lines indicate the same \ion{H}{1} extents as in the correlation map.  Black or gray 
signal below the dashed line at -2.568 indicates concentrations of correlation signal in 
Figure~\ref{IC4182_LBT_corr} that are 
very unlikely ($>3\sigma$) to be generated by random noise.}}
\label{IC4182_LBT_corr_probs}
\end{figure}

\begin{figure} 
\center{
\includegraphics[scale=0.4,angle=90]{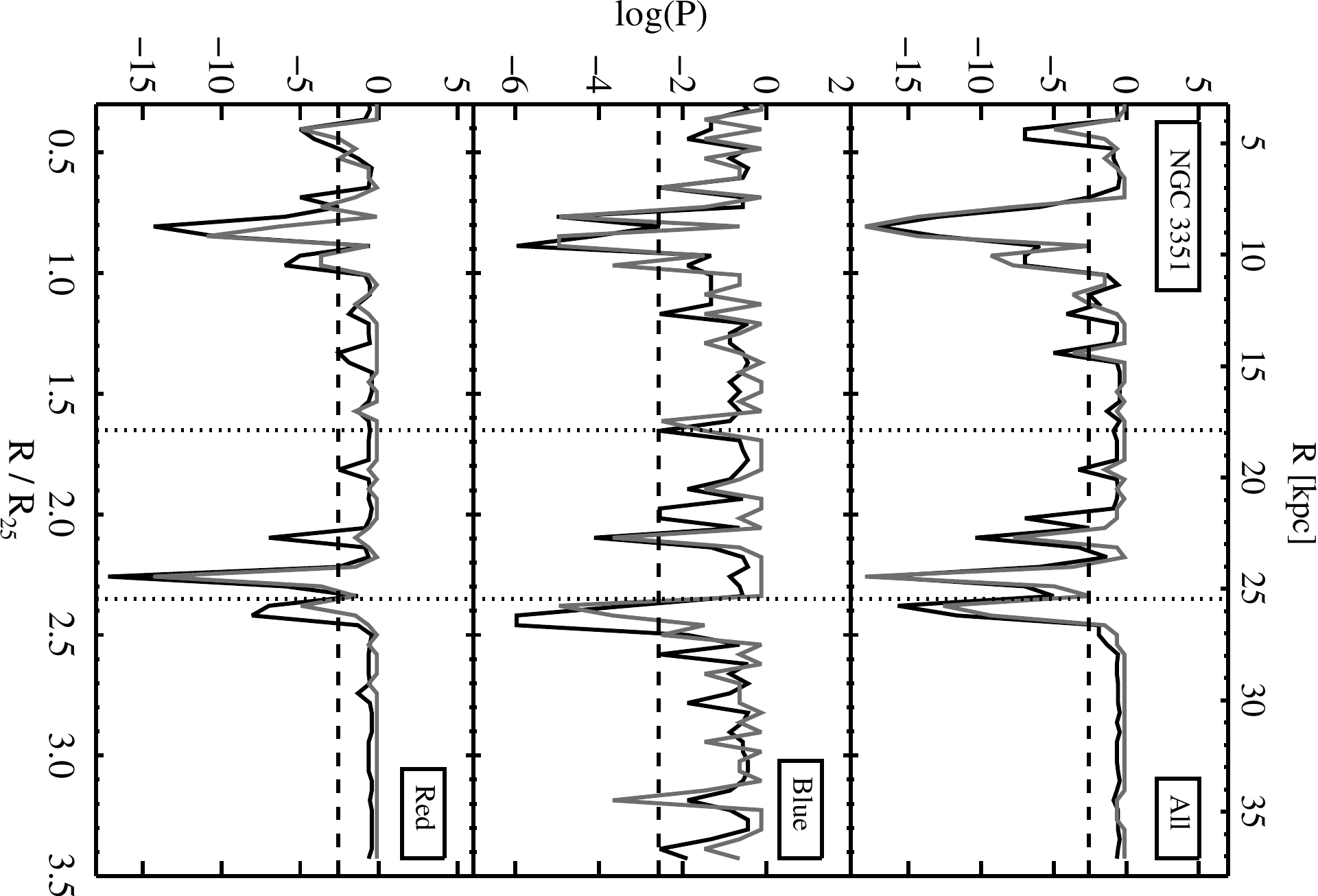}
\caption{Same as Figure 16  but for NGC 3351. }}
\label{N3351_LBT_corr_probs}
\end{figure}

\begin{figure} 
\center{
\includegraphics[scale=0.4,angle=90]{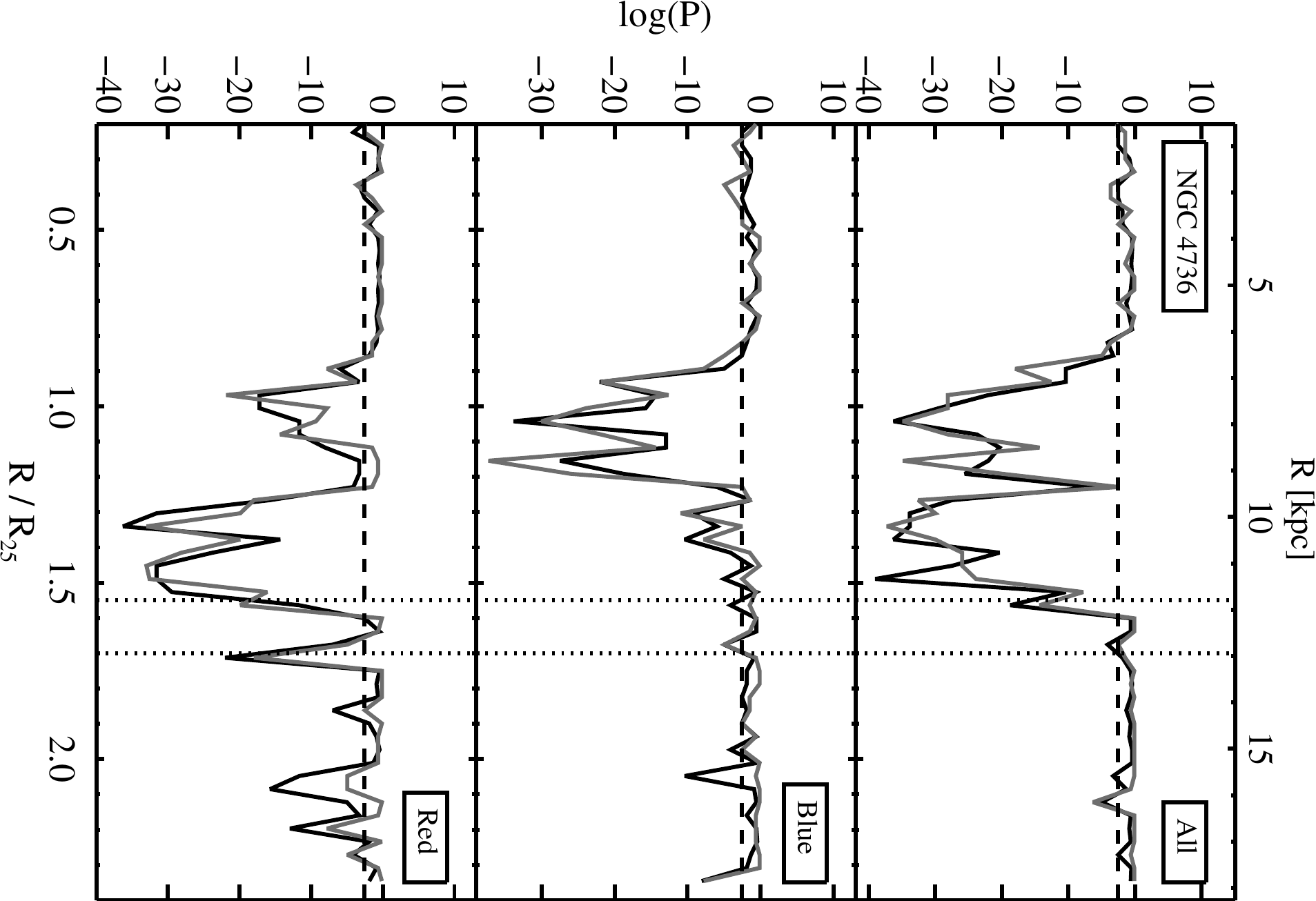}
\caption{Same as Figure 16 but for NGC 4736. }}
\label{M94_LBT_corr_probs}
\end{figure}

\begin{figure} 
\center{
\includegraphics[scale=0.4,angle=90]{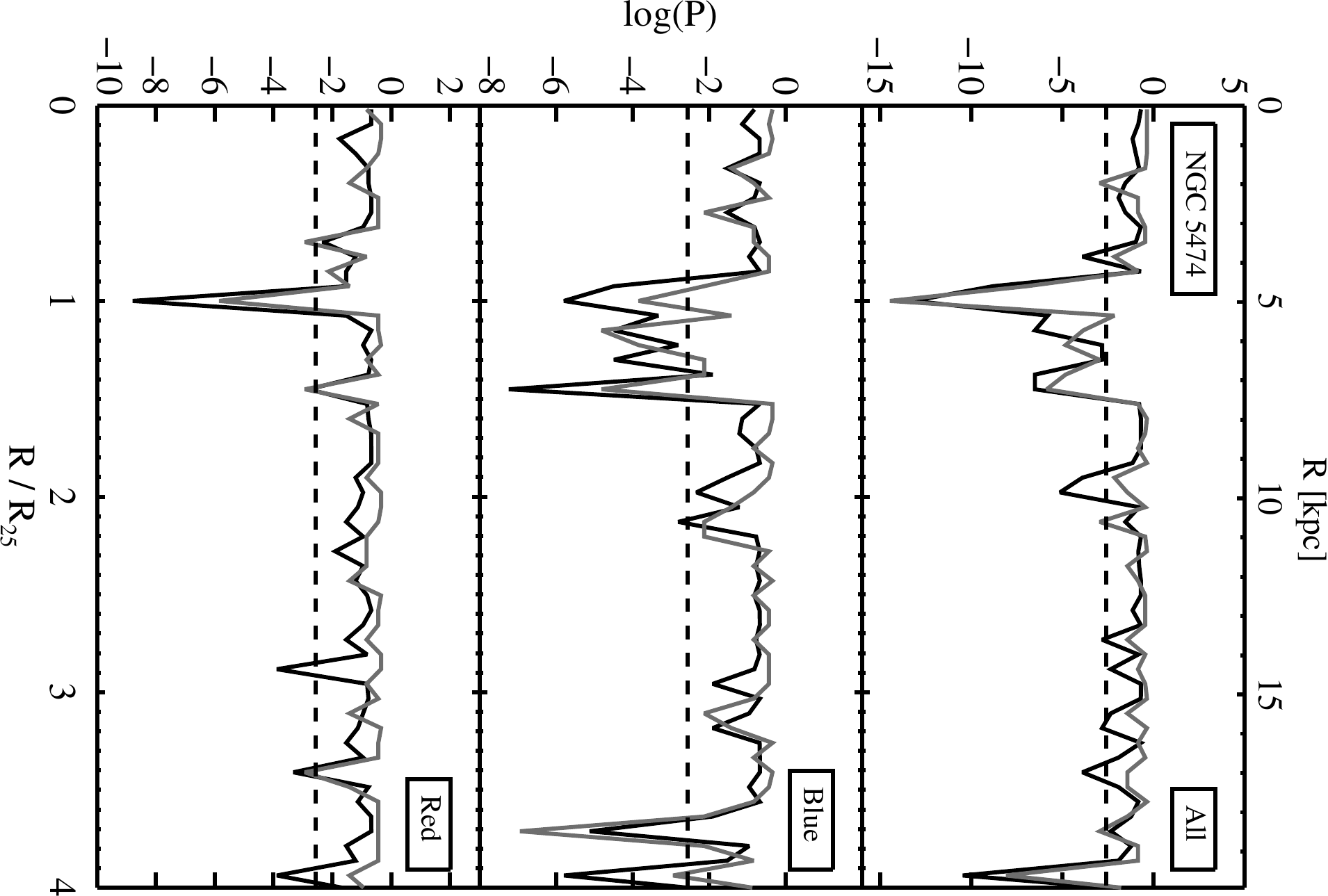}
\caption{Same as Figure 16 but for NGC 5474. }}
\label{N5474_LBT_corr_probs}
\end{figure}

When possible, we combine the GALEX catalogs 
from shallow ($t_{exp} < 1000$ seconds) and deep exposures 
to include sources both near and far from the bright inner galactic disk (regions around the galaxies were masked in the catalog generation for the deep GALEX exposures and so the shallow exposures 
help fill in the source distribution).  We then cut 
the combined GALEX source lists to match the `blue' sample of \cite{Zaritsky07} with FUV$-$NUV $< 1$ and 
NUV $< 25$, corresponding to clusters younger than $\sim 360$ Myr.  This cut removes sources 
with detections in only one GALEX band and retains the most reliable knots.  We 
reduce contamination further by only retaining knots that match 
sources in our LBT catalogs (to within $3^{\prime\prime}$, roughly half a GALEX 
resolution element). The large majority, $> 90\%$, of the GALEX knots positionally match an LBT source.

Before proceeding to the correlation analysis,  we 
compare the number of GALEX knots around each galaxy 
with those remaining in the background-subtracted Hess diagrams (Figure 3 --- 9) and present 
the results for two outer disk annuli in each low-inclination 
galaxy, between $1.0-1.5 R_{25}$ and $1.5-2.0 R_{25}$, in Table \ref{tab:GALEX}.  
Unfortunately, the inner annulus is problematic because the GALEX catalogs near the disk
are made from shallow exposures. 
In the $1.5-2.0 R_{25}$ annulus, we find that GALEX knots make up between 10\%-40\% of LBT knots, 
or 27\% on average.  For a population of clusters that form continually, this ratio should simply reflect the
timescales over which the clusters would be detected in GALEX versus LBT imaging, if cluster dissolution affects
both samples similarly. Since dissolution is expect to proceed very rapidly \citep[few million year timescale; see][]{Fall05}  both the GALEX and LBT samples, unlike H$\alpha$ samples, 
are dominated by clusters that `survive'. The average age of
the GALEX population is $\sim 180$ Myr 
(assuming a uniform distribution of ages between $0 - 360$ Myr, where the upper bound is set by
the UV color selection), while that of the LBT clusters is $\sim$ 500 Myr  (assuming a uniform distribution between $0 - 1$ Gyr, where the upper limit is again given by the color and magnitude selection). The ratio of these mean ages leads to an 
estimated fraction of 36\%, sufficiently close to our detected average of 27\% that we cannot rule out
the hypothesis of continual cluster formation.  A lower fraction implies a 
higher cluster formation rate in the past, while a higher fraction implies a lower formation 
rate.  The observed scatter between 10\% - 40\% suggests that factor of two variations in the cluster formation rate over these timescales are likely, but that the rates do not change by orders of magnitude over the previous Gyr when integrated over timescales of $\sim200$ Myr (roughly the resolution limit of our crude age
estimates).

\begin{figure} 
\includegraphics[scale=0.4,angle=90]{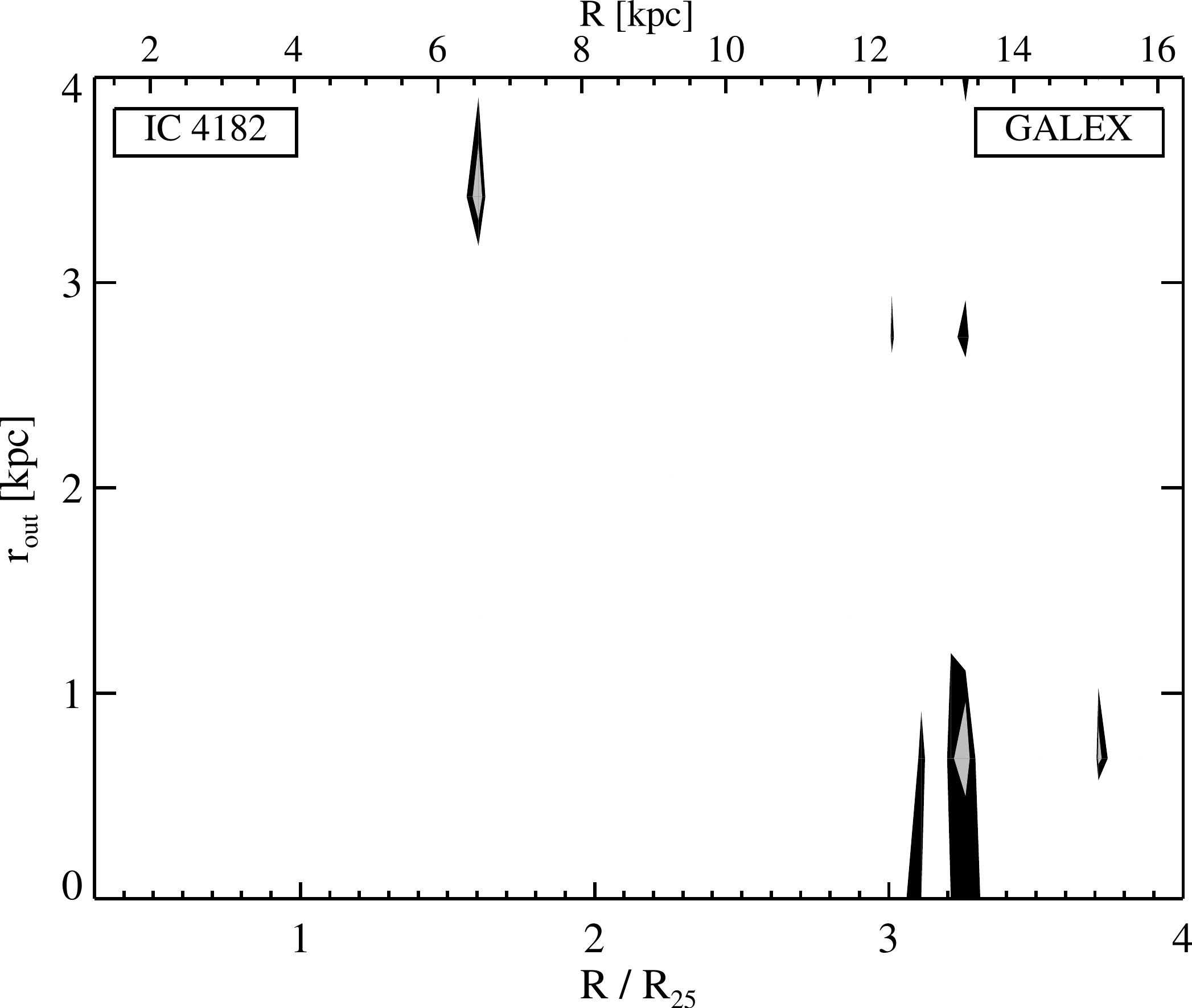}
\caption{Restricted three-point correlation map similar to Figure~\ref{IC4182_LBT_corr} but here using 171 
GALEX sources around IC 4182.  The statistics are limited in comparison to the LBT self-clustering analysis by the relative shallower depth and poorer resolution of GALEX.}
\label{IC4182_GALEX_corr}
\end{figure}

\begin{figure} 
\includegraphics[scale=0.4,angle=90]{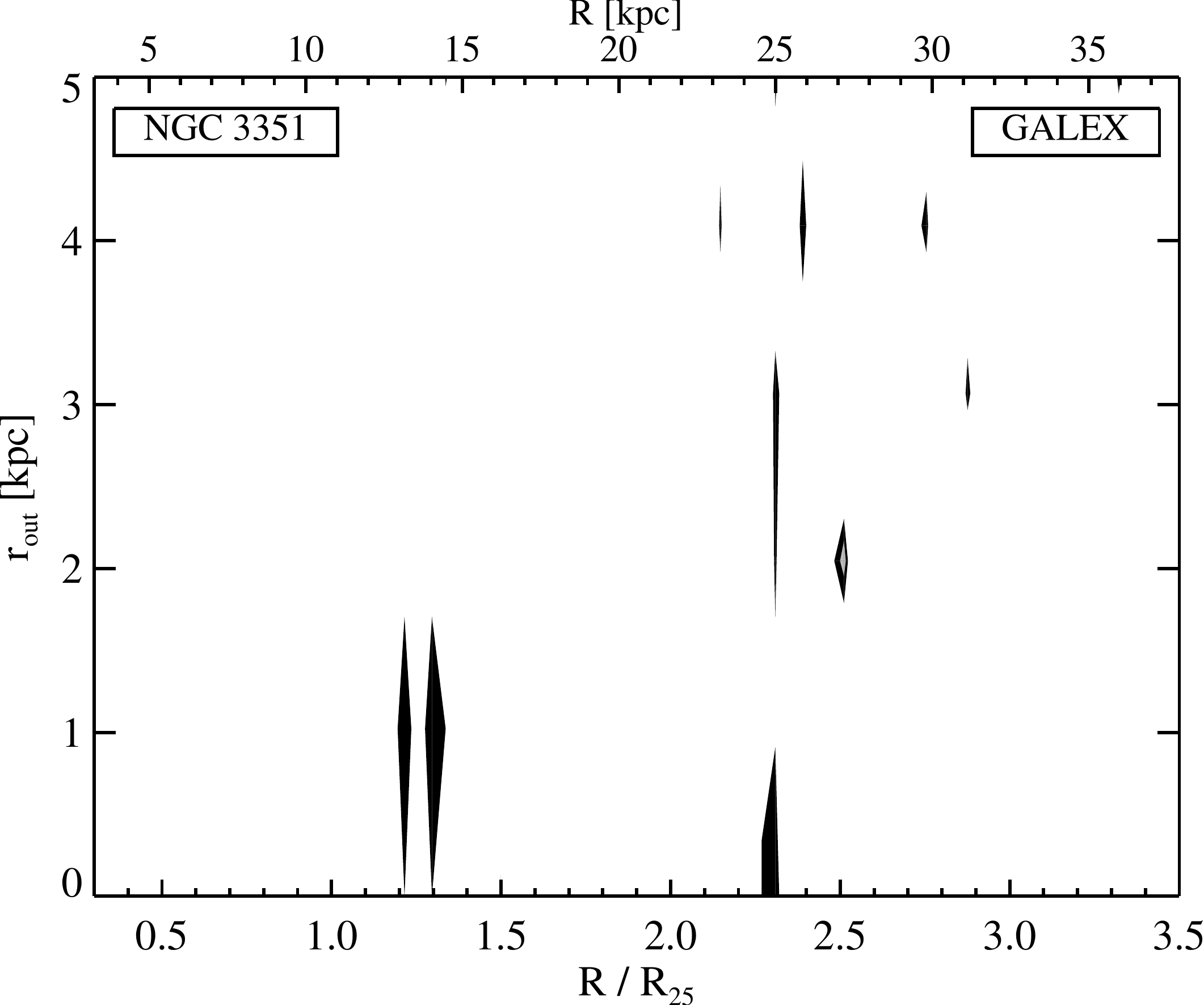}
\caption{Same as Figure~\ref{IC4182_GALEX_corr} but from 179 GALEX sources around NGC 3351.}
\label{N3351_GALEX_corr}
\end{figure}

\begin{figure} 
\includegraphics[scale=0.4,angle=90]{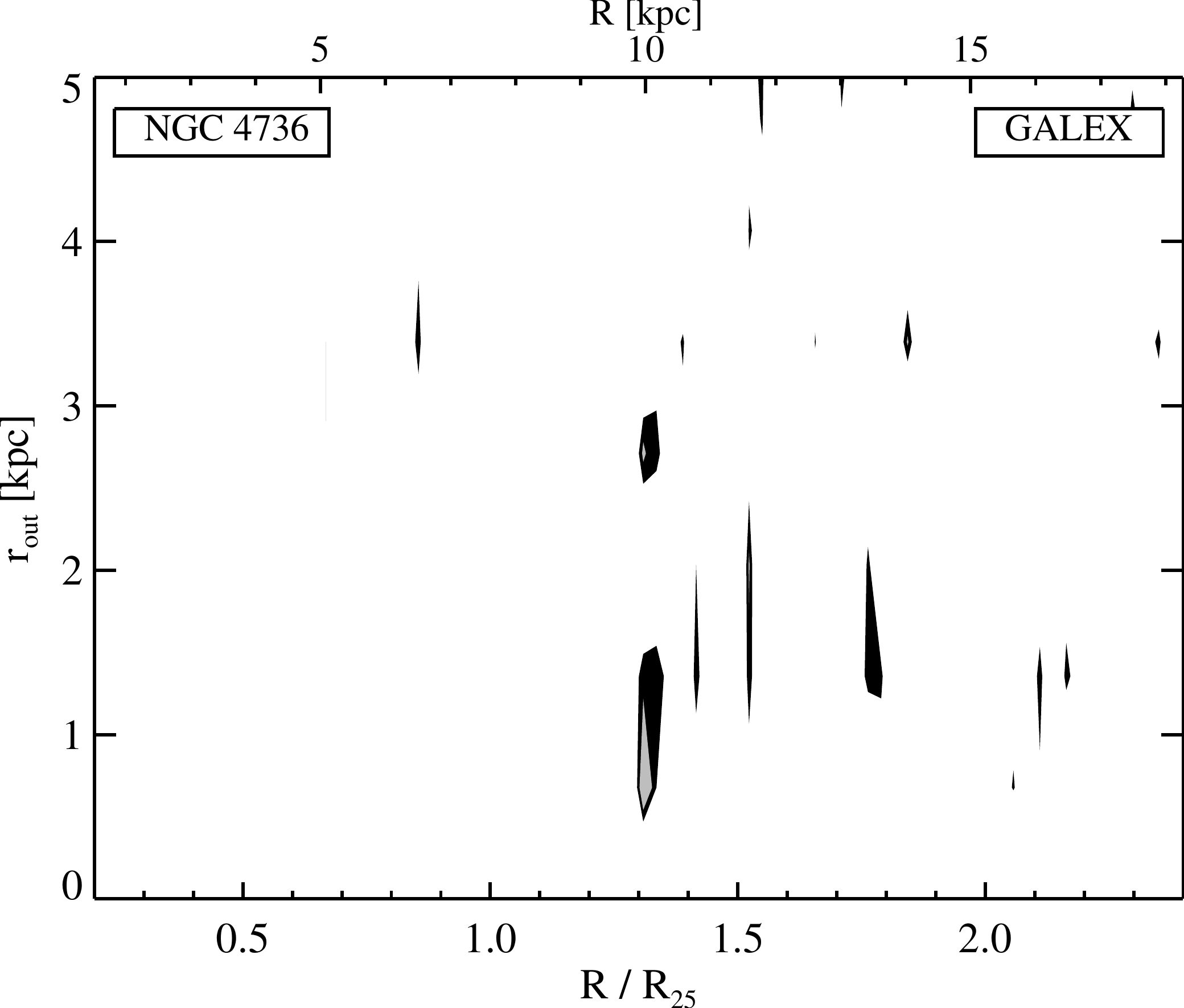}
\caption{Same as Figure~\ref{IC4182_GALEX_corr} but from 398 GALEX sources around NGC 4736. }
\label{M94_GALEX_corr}
\end{figure}

\begin{figure} 
\includegraphics[scale=0.4,angle=90]{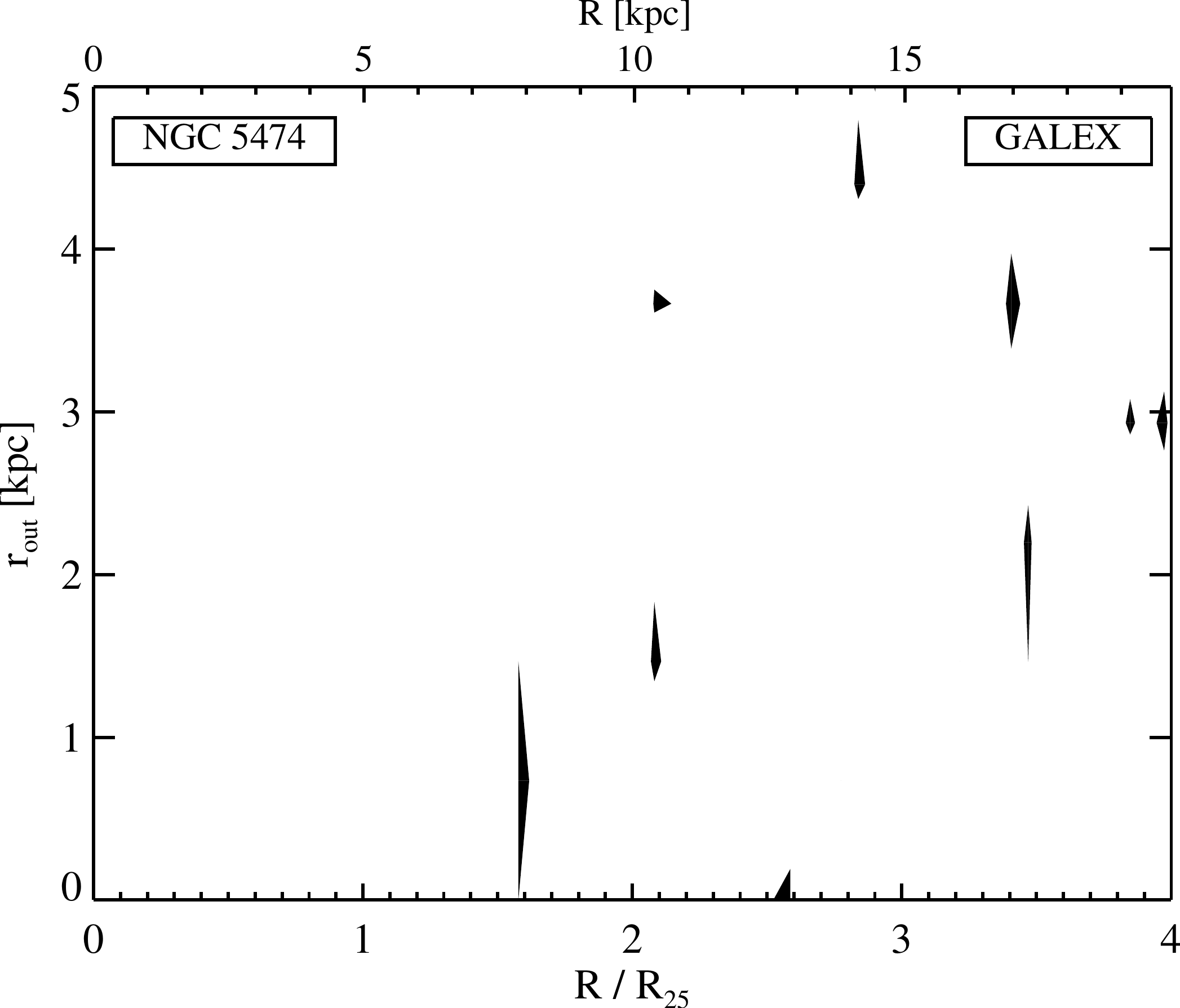}
\caption{Same as Figure~\ref{IC4182_GALEX_corr} but from 174 GALEX sources around NGC 5474. }
\label{N5474_GALEX_corr}
\end{figure}

We present the UV knot self-clustering maps from our 
final GALEX source catalogs in Figures~\ref{IC4182_GALEX_corr} to  \ref{N5474_GALEX_corr}.  Unfortunately, the number of available GALEX sources is low; 171 were used 
around IC4182, 179 around NGC 3351, 398 around NGC 4736, and 174 around NGC 5474.  
Of these, NGC 4736 and NGC 5474 were included in the \cite{Zaritsky07} study. They found
an excess of knots in both cases, although the probability of the excess being random is
not exceedingly small, $4\%$ for NGC 4736, so the results based solely on the radial distribution
of GALEX knots are marginal. Interestingly, the LBT results for NGC 4736 are the strongest in our optical sample, and so
show the value of going to deeper, redder samples in uncovering outer disk populations.

\subsection{Cross-correlation of LBT Knots and Neutral Gas Disks}

We now compare the distribution of our LBT sources to the neutral gas distribution.  
This cross-correlation provides another way to constrain the extent of the 
stellar disks, particularly in relation to the known component that stretches to 
the farthest radii. Any association between the \ion{H}{1} structure 
and knots is also key for understanding how star formation arises in this 
environment.

Once again, because we perform an angular cross correlation analysis, we are restricted to our 
low-inclination sample (IC 4182, NGC 3351, NGC 4736, and NGC 5474). A further limitation is that
we do not have a suitable \ion{H}{1} map of 
NGC 5474.  The integrated \ion{H}{1} (moment 0) maps for NGC 3351 and NGC 4736 come from 
The \ion{H}{1} Nearby Galaxy Survey \citep[THINGS; ][]{Walter08} and for IC 4182 from the \ion{H}{1} map of 
from the Westerbork \ion{H}{1} Survey of Spiral and Irregular Galaxies 
\citep[WHISP; ][]{Swaters02}.  
We estimate the extent of the gas in the same manner as described by \cite{HF09}, by 
examining histograms of the number of \ion{H}{1} pixels per kpc$^2$ above a 
certain $N$(\ion{H}{1}) threshold, as a function of $R$ in $0.025 R_{25}$-wide elliptical annuli. 

The first threshold we consider is the noise level of the \ion{H}{1} maps (see below), 
to determine the maximum detected extent of the gas in each of the three galaxies.  
The `noise level' of the final \ion{H}{1} maps results from already having cleaned 
the maps, only keeping pixels where two or more adjacent channels show significant signal ($2\sigma$) in the integrated moment 0 map (so the `noise level' here is really more significant that the `2$\sigma$' noise level in the raw data).
The neutral gas around IC 4182 extends to $\sim2.25 R_{25}$ around IC 4182, to $\sim2.4 R_{25}$ around NGC 3351, 
and to $\sim1.8 R_{25}$ around NGC 4736.  Figures~\ref{IC4182_LBT_HI_corr} --- \ref{M94_LBT_HI_corr} 
show the restricted three-point cross-correlation between the LBT knots and the \ion{H}{1} pixels 
lying above the $N$(\ion{H}{1}) threshold (listed in the legend at the bottom right of each panel).  
The top panels show results using the `All'  sample while the 
lower panels give the results from using the `Blue' and `Red' samples
(middle and bottom panel, respectively).  Black and gray 
show areas where signal is detected at $>95\%$ and $>99\%$ significance, respectively.  
The dotted lines bracket the radial extent of the \ion{H}{1} pixels used.  
The Figures show that LBT knots correlate with \ion{H}{1} features out to, and beyond, the observed edges  of the gas disks. As \cite{HF09} discussed, the vertical bands of signal highlight likely spiral arm structures.

To determine what \ion{H}{1} density might be the best tracer of knot formation, 
we focus on our `Blue' LBT knot sample and calculate correlations relative to the \ion{H}{1} distribution as defined by different column density thresholds.
Figures~\ref{IC4182_LBT_HI_BLUE_corr} --- \ref{M94_LBT_HI_BLUE_corr}  contain results
specifically for $N$(\ion{H}{1}) $> 1.0\sci{20} \cm{-2}$ (middle), and $N$(\ion{H}{1}) $> 2.0\sci{20} \cm{-2}$ (bottom).  The highest threshold, 
$N$(\ion{H}{1}) $> 2.0\sci{20} \cm{-2}$, corresponds to where one finds damped Lyman-$\alpha$ (DLA) absorption, and so, distinguishes regions that contain predominantly neutral or ionized 
gas \citep[see][and references therein]{Wolfe05}.  
The $N$({\ion{H}{1}) $= 2\sci{20} \cm{-2}$ threshold highlights the edge 
of the dominant reservoir of neutral gas.  The 
$N$(\ion{H}{1}) $= 1.0\sci{20} \cm{-2}$ threshold was determined by eye to be the lowest density value at
which the contours closely trace the distribution of the `Blue' knots.  
The radial extent of the \ion{H}{1} decreases as the column density threshold 
increases, though as the density threshold is increased we are more likely selecting \ion{H}{1} structures 
that could host cluster formation --- unless the gas has been consumed to make the clusters or dispersed after cluster formation. Broadly, we find correlations between the knots and the gas at all gas densities and at all radii. We discuss individual cases and reach conclusions next.

\begin{figure} 
\includegraphics[scale=0.7,angle=90]{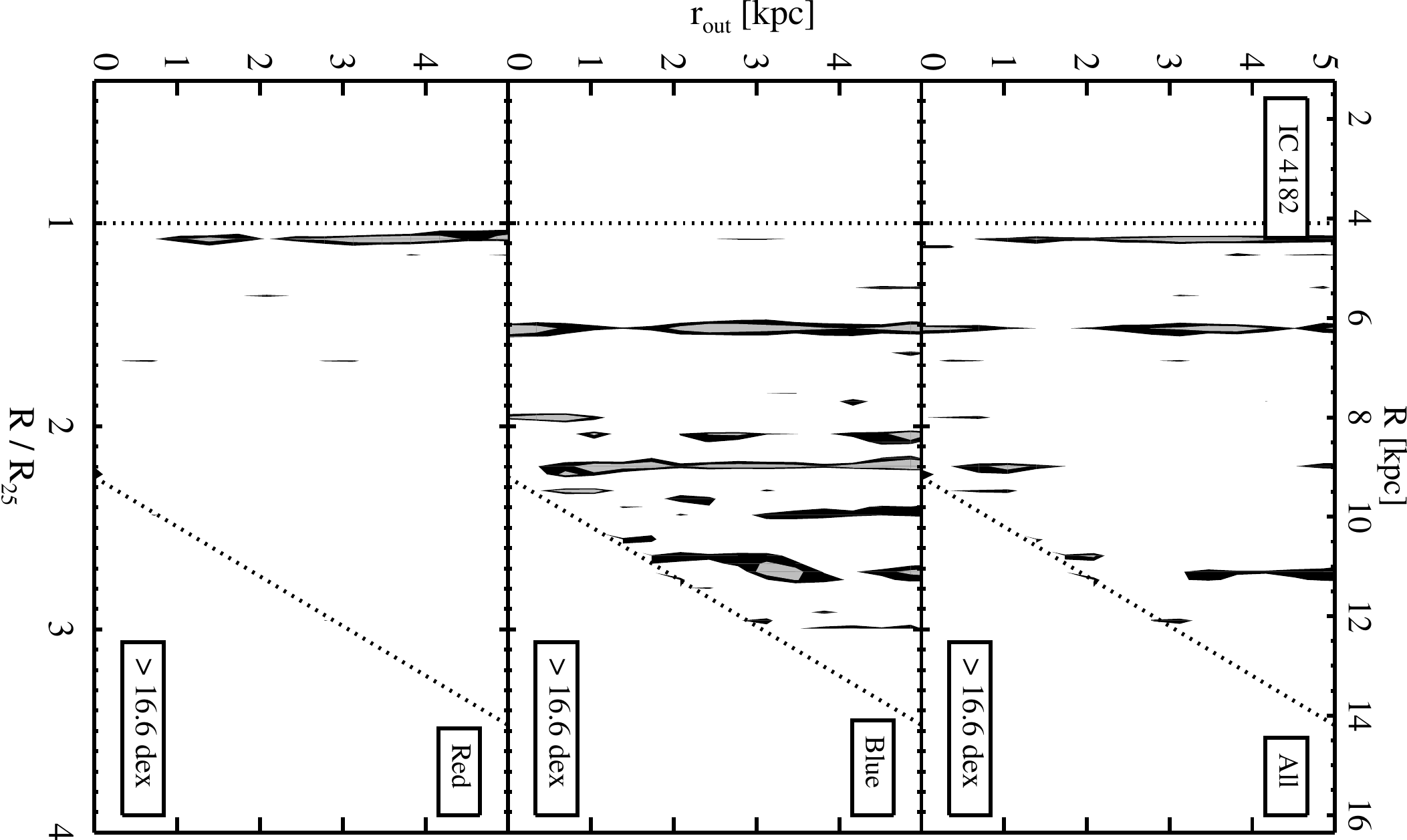}
\caption
{Restricted three-point cross-correlation maps of 
LBT-detected knots around IC 4182 and \ion{H}{1}\ pixels with $N$(\ion{H}{1}) above the noise level 
of the integrated \ion{H}{1} map ($N$(\ion{H}{1}) $> 4.0\sci{16} \cm{-2}$ here).  
The dotted lines bracket the radial extent of the \ion{H}{1} pixels used; because no \ion{H}{1} pixels 
exist beyond the dotted line on the right, we will not see signal at low $r_{out}$ 
beyond this furthest \ion{H}{1} radius (this explains the diagonal nature of signal at the largest $R$). 
Only knots between $1.0-3.0 R_{25}$ were used.  }
\label{IC4182_LBT_HI_corr}
\end{figure}

\begin{figure} 
\includegraphics[scale=0.7,angle=90]{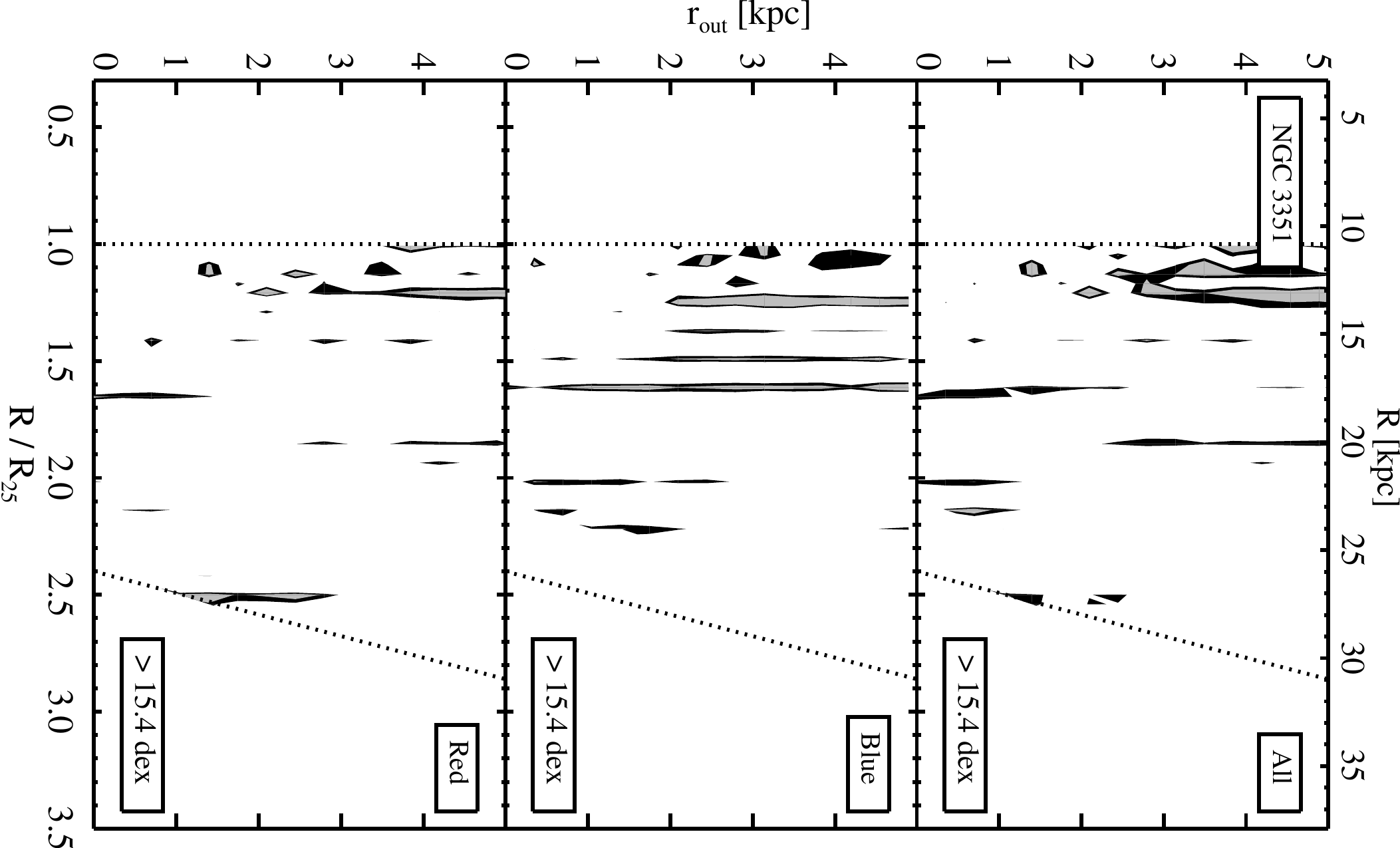}
\caption{Same as Figure~\ref{IC4182_LBT_HI_corr} but for NGC 3351.   Only sources between $1.0-3.5 R_{25}$ were used.}
\label{N3351_LBT_HI_corr}
\end{figure}

\begin{figure} 
\includegraphics[scale=0.7,angle=90]{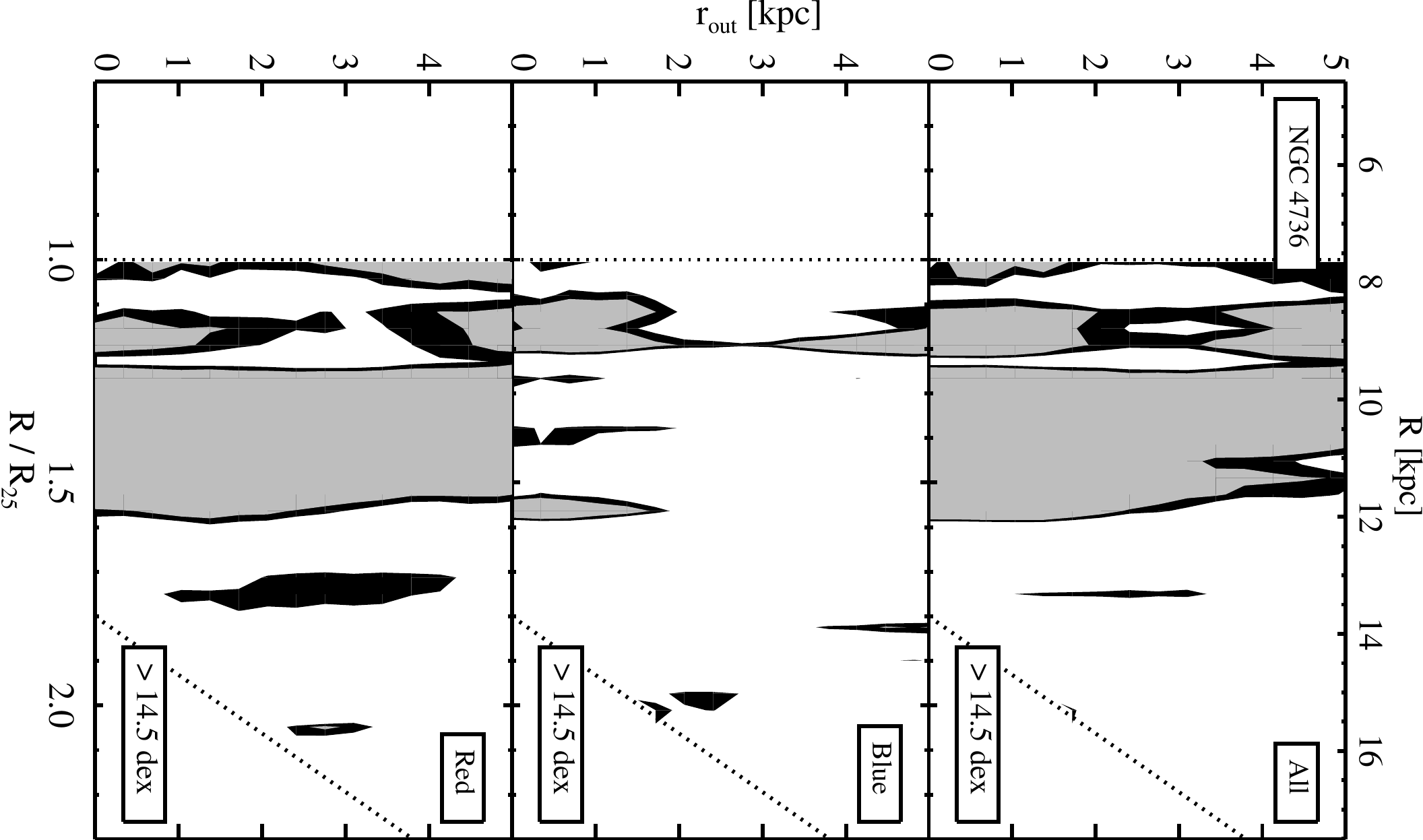}
\caption{Same as Figure~\ref{IC4182_LBT_HI_corr} but for NGC 4736.  Only sources between $1.0-2.2 R_{25}$ were used.}
\label{M94_LBT_HI_corr}
\end{figure}

\begin{figure} 
\includegraphics[scale=0.7,angle=90]{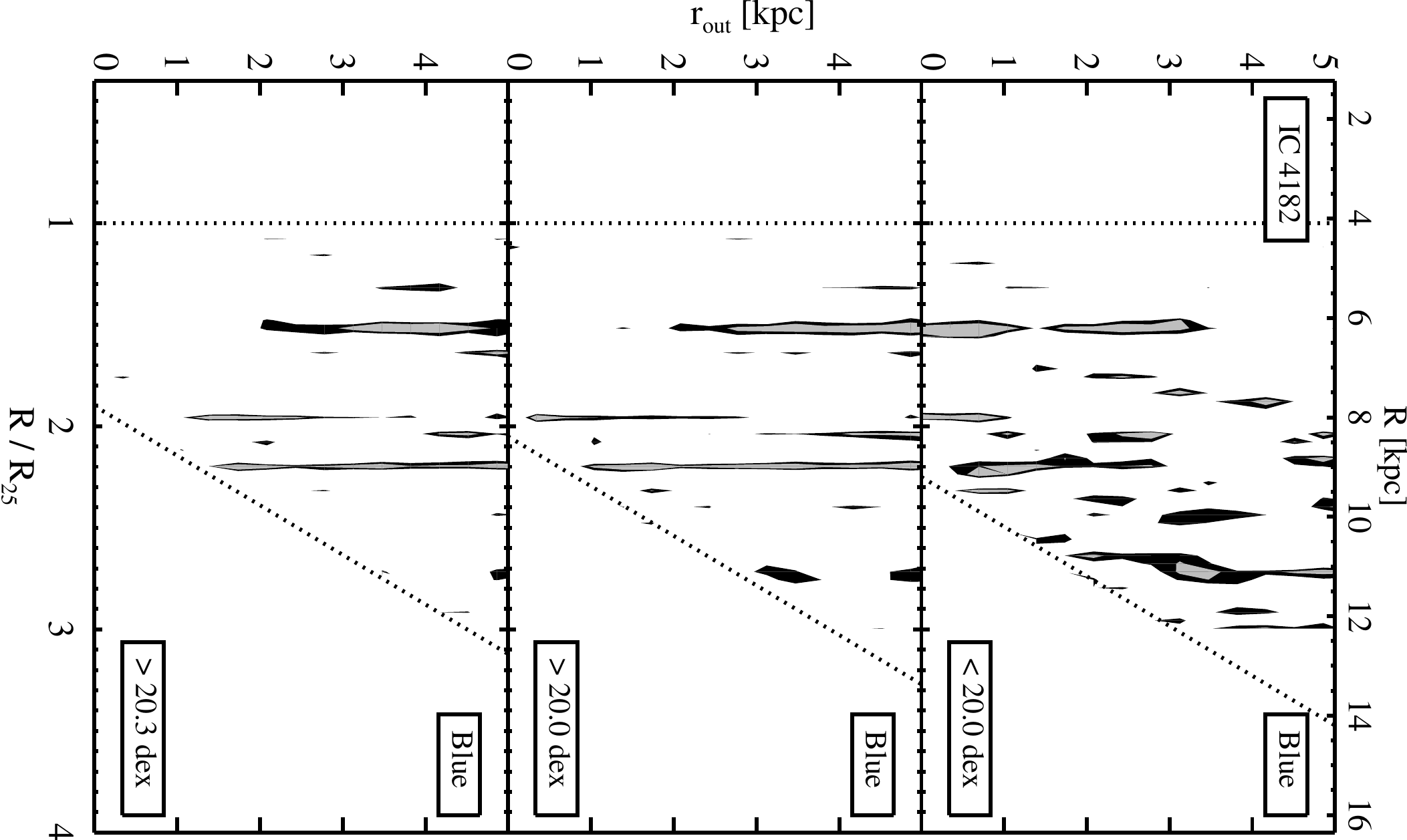}
\caption{Same as Figure~\ref{IC4182_LBT_HI_corr}, but only for the Blue sample of IC 4182 and for 
different $N$(\ion{H}{1}) thresholds.   }
\label{IC4182_LBT_HI_BLUE_corr}
\end{figure}

\begin{figure} 
\includegraphics[scale=0.7,angle=90]{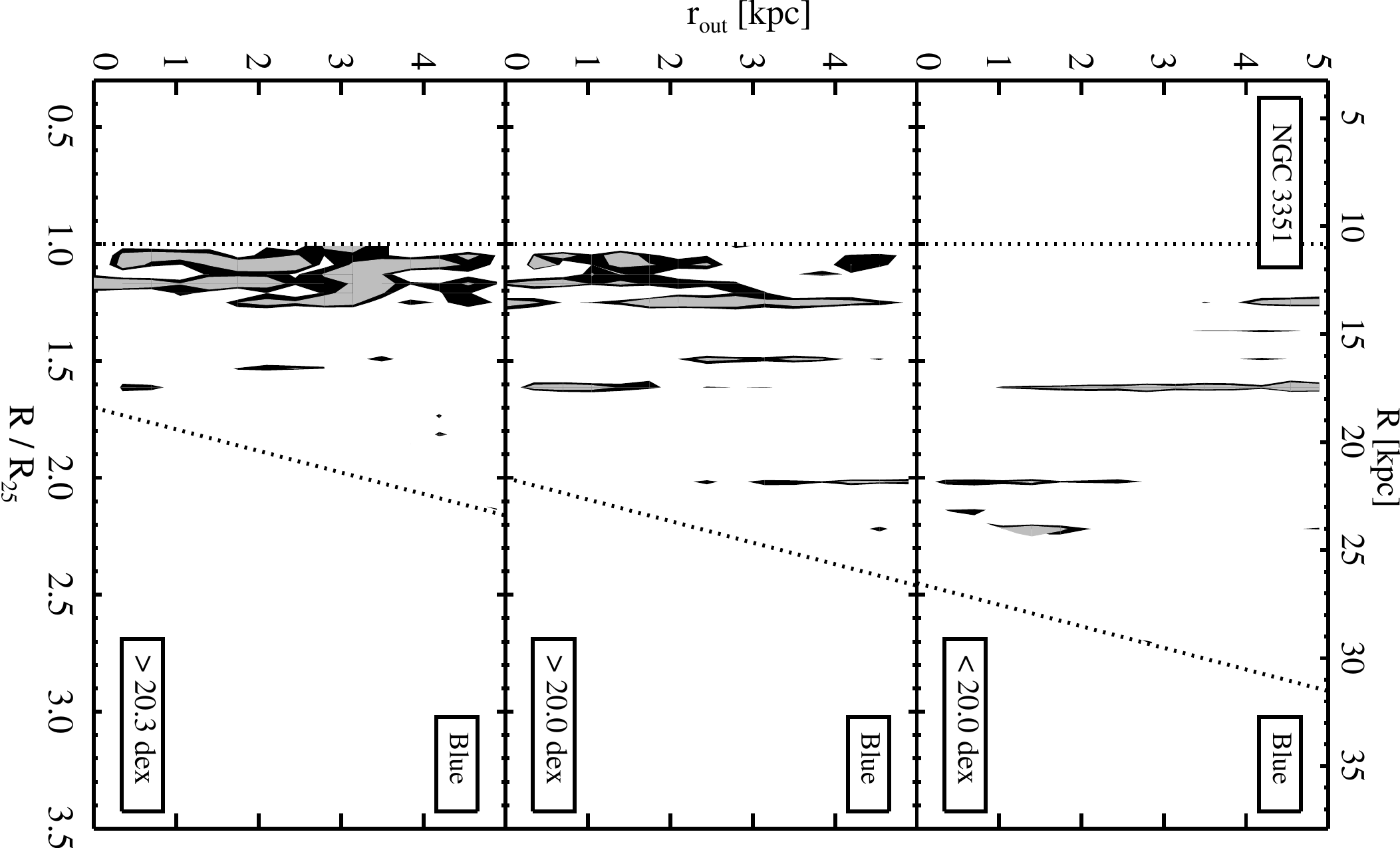}
\caption{Same as Figure~\ref{IC4182_LBT_HI_BLUE_corr} but for NGC 3351. }
\label{N3351_LBT_HI_BLUE_corr}
\end{figure}

\begin{figure} 
\includegraphics[scale=0.7,angle=90]{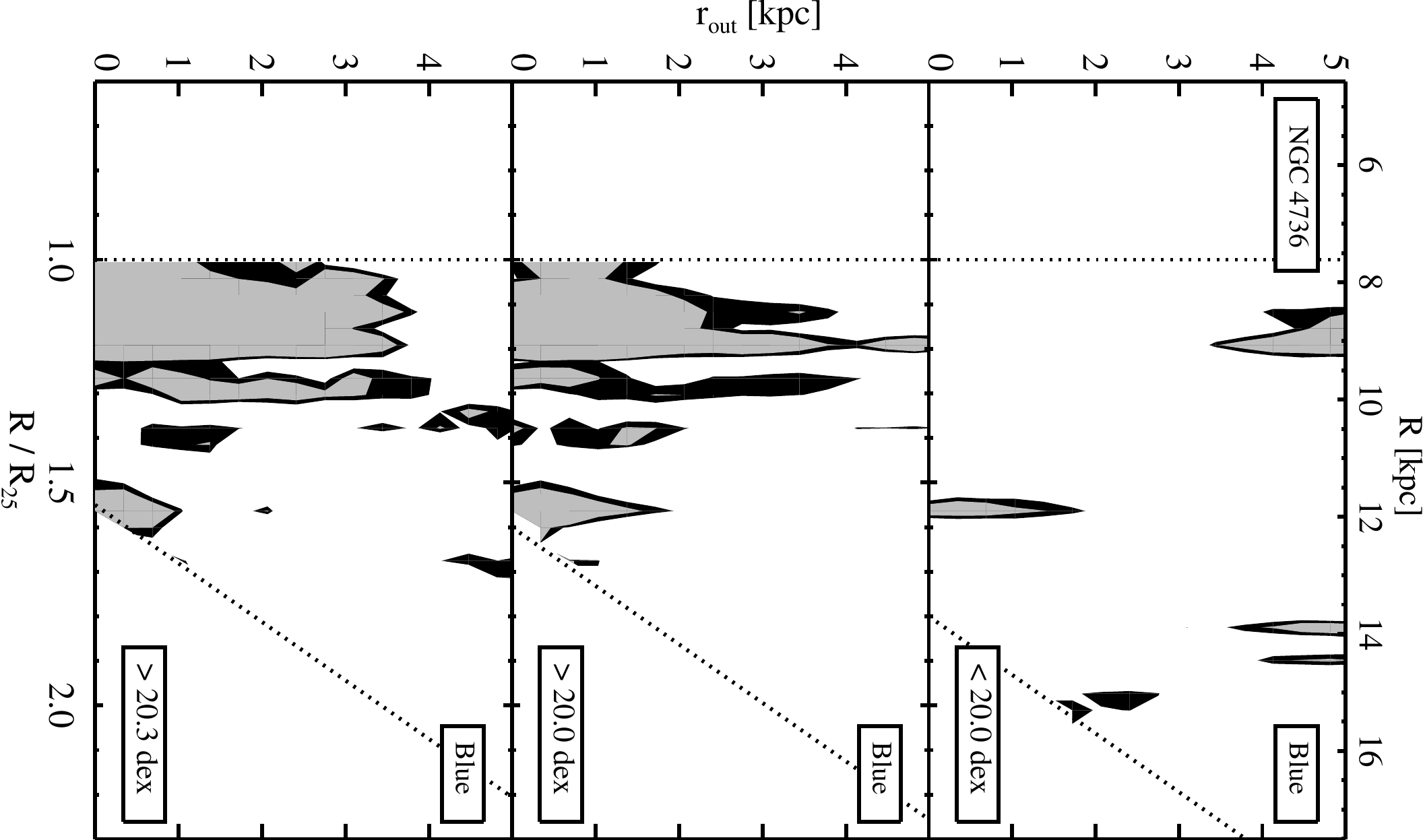}
\caption{Same as Figure~\ref{IC4182_LBT_HI_BLUE_corr} but for NGC 4736. }
\label{M94_LBT_HI_BLUE_corr}
\end{figure}

\section{Results for Individual Galaxies}

A few general comments before we discuss the galaxies individually.
In all four LBT correlation maps (Figures~\ref{IC4182_LBT_corr} ---  \ref{N5474_LBT_corr}) 
there is significant signal at very 
large radii, in three cases extending beyond the \ion{H}{1} (we do not have \ion{H}{1} data 
for the fourth galaxy, NGC 5474).  Though we are somewhat skeptical of correlation signal at
such large radii, where the background becomes relatively more important, these detections
may point to stars formed beyond the current edge of the observable \ion{H}{1} disk.
We examined the galaxy images and, with one exception that we discuss below, 
could not visually identify groupings of LBT knots at radii beyond 
the \ion{H}{1} extent.  However, the knots can be very difficult to find when they are far from the disk, not 
strongly clustered, and faint, which is of course why we devised the restricted 3-pt correlation analysis.  When we compare  correlation maps made using the masked and unmasked source 
distributions (masked to limit the catalog problems near bright stars, as mentioned in \S2), much of 
the signal beyond the \ion{H}{1} changes significantly in character, suggesting it might be artificial.  
We hesitate to increase the masking beyond what we already chose because of the limitation in the number of sources near the 
mosaic edges.  Nevertheless, 
although some of the most distant signal may be artificial, 
we will demonstrate that some of the features are real and that there are clusters out to at least $\sim 3 R_{25}$
and beyond the edge of the \ion{H}{1} disk.
We now discuss the results for each galaxy.

\subsubsection{IC 4182}

We find the bulk of the signal contained within $\sim R_{25}$ in the LBT knot 
correlation map, although there is a great deal of structure in the Blue knot correlation map at larger
radii, and some concentration of signal toward low $r_{out}$ in the Red one (Figure \ref{IC4182_LBT_corr}).
Working our way out in radius, the first tantalizing detection in the 2-D plot is in the Red knots
at about $\sim 1.5 R_{25}$. However, this turns out to not be statistically significant (Figure \ref{IC4182_LBT_corr_probs}) and
is also likely to be the
remnant of structure caused by a bright star at that radius.
However, strong signal exists near $1.5 R_{25}$ in the All and Blue panels 
of the LBT knot - \ion{H}{1} 
cross-correlation plot (Figure~\ref{IC4182_LBT_HI_corr}). 
The lack of a self-clustering signal among the Blue sample is not counter-evidence, 
just evidence for a lack of strong self-clustering among such clusters.
Interestingly, the Blue LBT knot - \ion{H}{1} cross correlation at $1.5 R_{25}$ is most closely 
related to the lower-density gas, $N$(\ion{H}{1}) $ < 20.0$ dex (Figure \ref{IC4182_LBT_HI_BLUE_corr}), suggesting that these knots have either
moved away from the regions of highest gas density or affected their nearby environs.

Further out in $R$, the Blue panel of the LBT knot probability plot 
(Figure 16) 
shows more noisy, scattered signal beyond the edge of the \ion{H}{1} than 
do the All or Red panels.  
This `noisy' Blue signal in the LBT 
correlation map between $2-3 R_{25}$ is statistically significant in the probability plot and 
coincides with bands of Blue signal in the LBT knot - \ion{H}{1} cross-correlation 
plots, suggesting that Blue clusters a few kpc beyond the edge of the gas 
are nevertheless strongly associated with the outer disk gas structure.  This result suggests that the knots 
were either born from gas that was once part of the disk, but no longer exists, or that they have drifted
somewhat in radius from their birthsites. A drift velocity of just 6 \kms\ could transport a cluster a few kpc 
in 500 Myr.
The GALEX correlation plot (Figure \ref{IC4182_GALEX_corr}) does not contain signal at these radii, which suggests that these clusters are somewhat older than a few hundred Myr. This scenario is consistent with
the inferences discussed above that the clusters are not as strongly correlated with the highest density gas and that they may have drifted in radius.

Finally, all of the panels in the LBT knot correlation probability plot (Figure 16) show significant excess signal at $\sim 3.3 R_{25}$, 
far outside the edge of the gas disk (the gas extends to $2.2 R_{25}$).  
This feature is especially interesting because it corresponds to strong signal at the same radius
in the GALEX knot self-correlation map (Figure~\ref{IC4182_GALEX_corr}).
Examining the GALEX image, we visually identify
a clump of sources at that 
radius.  If these distant sources are in fact outer-disk clusters, rather than a grouping of UV-bright background 
objects or stellar clusters in a satellite galaxy, they are very interesting objects for further study.  
H$\alpha$ imaging, and subsequent spectroscopy would be particularly valuable.

\subsubsection{NGC 3351}

The bulk of the signal in the LBT knot correlation map and corresponding probability plot
(Figures~\ref{N3351_LBT_corr} and 17) is once again contained
within $\sim R_{25}$, demonstrating that the optical radius does demarcate a real qualitative 
change in the nature of star formation. Again moving out in radius, the most significant signal
comes in at around 2.4$R_{25}$ in the All and Red panels, and to a lesser degree in the Blue
panel. These clusters are also seen in the GALEX correlations (Figure \ref{N3351_GALEX_corr}).
Evidence for correlation between this population and the \ion{H}{1} is weaker, although there
is some signal near $2.2 R_{25}$ in the Blue panel and $2.5 R_{25}$ in the Red panel 
(Figure~\ref{N3351_LBT_HI_corr}).
This signal occurs right at the edge of the \ion{H}{1} disk.

In addition to the clusters seen at the periphery of the \ion{H}{1}, there is also signal 
in the LBT knot - \ion{H}{1} cross-correlation map form $R_{25}$ out to
$1.6 R_{25}$.
From the Blue LBT knot - \ion{H}{1} cross-correlation map for different $N$(\ion{H}{1}) thresholds
(Figure~\ref{N3351_LBT_HI_BLUE_corr}), we conclude that 
the knots trace the denser gas out to 
$\sim1.6 R_{25}$ (and trace spiral structure out to $1.2 R_{25}$). As in
IC 4182, the bands of correlations between the Blue knots and \ion{H}{1} come at semi-regular radial spacings, suggestive of
multi-armed spirals with small pitch angles or rings.

\subsubsection{NGC 4736}

As we have discussed before, NGC 4736 is by far the richest of our galaxies in
outer disk structure and may therefore not be representative. Nevertheless, we
see many of the same general patterns. 
The LBT knot correlation map and corresponding probability plot for 
NGC 4736 (Figures~\ref{M94_LBT_corr} and 18) 
show signal extending to at near the limit of the HI disk ($\sim1.6 R_{25}$). 
This galaxy has an 
obvious overdensity of knots in the bright `ring' surrounding it.  
\cite{Trujillo09} found that this 
ring is actually a complex structure of wound spiral arms (perhaps caused by secular processes in the inner 
disk), which perhaps are a much stronger version of the low pitch angle arms we discussed in the context of IC 4182 and NGC 3351. The knots toward the inner edge of the outer disk are associated with higher
density \ion{H}{1}, as is also the case for NGC 3351.
We are somewhat
skeptical of excess signal beyond the \ion{H}{1} extent of this galaxy (signal at $R \sim 2 R_{25}$) because it is coincident 
with artificial signal that faded when we applied additional bright star masking, though there is 
some signal in the \ion{H}{1} correlation maps at the same radius. The lack of correlation in the GALEX
data is not a concern because the signal in the LBT correlation maps comes primarily from the Red
sample.
 
\subsubsection{NGC 5474}

NGC 5474 is our lowest noise LBT knot correlation map and contains significant signal from $R_{25}$ to $\sim1.4 R_{25}$ (Figures \ref{N5474_LBT_corr} and 19).
Unfortunately, we do not have the same quality 
\ion{H}{1} data for NGC 5474 as for the other three galaxies. However, the 
GALEX knot correlation map contains some signal near $1.5 R_{25}$, though it is very weak (Figure \ref{N5474_GALEX_corr}).  There are two reasons why the signal might be weak. First, it is possible that the GALEX knots do not cluster very tightly (or do not cluster differently than the background). Second, it may be that the relative small number of GALEX knots limits the degree to which this restricted 3-pt correlation can result in a statistically meaningful detection. From looking
at the GALEX images and source distribution we conclude that this feature is real, but this weak signal (as well as the rather tame GALEX signal for NGC 4736) demonstrates how our optical work complements the
GALEX studies.

There are a few marginally significant detections at intermediate radii in the correlation probability plot (Figure 19). A potentially interesting detection is that in the All panel at a radius
slightly below 2$R_{25}$. This radius matches our estimate for the outer extent of \ion{H}{1} based on 
Figure 2 from \cite{Rownd94}. While the other three galaxies in our sample all show clear signs for
cluster formation near the periphery of their \ion{H}{1} distribution, we only have this questionable detection
for NGC 5474.

In the outer extremity of the correlation maps, $R \sim 3.5$ to 4 $R_{25}$, we again find strong signal in the three knot populations. There is no corresponding signature in the GALEX maps (although we saw above that the
GALEX detections can be weak) and, unfortunately, we do not have \ion{H}{1} maps 
of the required quality to help confirm these
features.  The feature near 4$R_{25}$ is likely residual signal from a few bright stars 
at that radius.

\section{Summary and Conclusions}

We summarize our results as follows:

\vspace{0.1in}

$\bullet$ All disk galaxies have a cluster population beyond $R_{25}$. The six for which we present background subtracted Hess diagrams show significant populations out to at least $1.5R_{25}$. The outer disk we studied similarly before \citep[NGC 3184;][]{HF09} also contains this population. Many of our galaxies show similar populations extending to $2 R_{25}$ and occasionally beyond. We attribute the larger fraction of galaxies with detected outer disk populations relative to previous studies \citep{Thilker07,Zaritsky07} to the superior mass and age sensitivity of the LBT data.

$\bullet$ Using the distribution of sources in the Hess diagrams, we infer that the typical detected cluster
has a mass of $\sim10^3 M_{\sun}$, is predominantly $< 1$ Gyr old, and as a population have an average formation
rate of at least $\sim$ one cluster every 2.5 Myr. The corresponding rate of stellar mass being formed in 
these clusters is $\sim0.004 M_{\sun}$ pc$^{-2}$ Gyr$^{-1}$ (assuming $R_{25} = 5$ kpc and $10^3 M_{\sun}$ clusters) for the area of the disk between $R_{25}$ and $1.5 R_{25}$. 
These estimates are rough, and may be systematically biased due to uncertainties in the modeling, stochastic effects in low mass cluster, and selection. The principal aim of this exercise was to demonstrate the plausibility of associating the detections with stellar clusters. Comparing the numbers
of sources identified to the corresponding numbers found using GALEX, which is sensitive only to 
younger clusters, we conclude that the formation rate of clusters at these radii is constant, to within
a factor of roughly 2, over the last Gyr.

$\bullet$  To further quantify the distribution of outer disk clusters, we construct restricted three-point correlation maps in our
four low-inclination galaxies. We confirm many of these detections using comparisons of GALEX detected populations and correlations with \ion{H}{1}.  
Again, we detect signal in all four, but our detections come in three
varieties. First, we generally find a population of clusters that extends modestly beyond the optical radii
(to between 1.3 and 1.5 $R_{25}$). Second, we find a population of clusters near the edge of the 
\ion{H}{1} distribution. Lastly, in all but NGC 3351, we find detections of clusters well beyond
the \ion{H}{1} edge. These last are the most difficult to confirm independently (they could either be
an unfortunate clustering of background sources or they could belong to 
satellite galaxies). 

$\bullet$  From the cross-correlation signal between our Blue LBT-detected knots and the \ion{H}{1} distribution 
we find two types of behavior. First, the knots near the optical edge of the disk are best traced by the higher density neutral gas, as one might expect in a steady state situation where density waves continually lead to the 
generation of new clusters. The pattern of the correlations, semi-regular bands in Figures \ref{IC4182_LBT_HI_corr} to \ref{M94_LBT_HI_corr}, also suggest spiral arms, in this case with small pitch angles.
Second, the knots farther out in the disk are most strongly correlated to the low density \ion{H}{1} gas,
suggesting that while some connection between star formation and fuel exists, the process is sufficiently
transient and/or disturbing that correlations with high density gas do not persist.

From these results, as well as those presented in previous studies of outer disks, we suggest that
outer disk cluster formation occurs in three modes. First, spiral waves from the inner disk continue
beyond the optical radius and trigger continual, but low level, cluster formation out to at most $\sim 1.5R_{25}$. We have presented some evidence for this mode of cluster formation, but it can also be clearly
seen in the H$\alpha$ images of NGC 628 \citep{Ferguson98} where the arms can be visually traced
beyond $R_{25}$. Second, a global mode, where clusters are formed throughout the disk, is triggered
by interactions. This mode is responsible for the rare and most dramatic examples of outer disk star formation, such as that seen in M83 \citep{Thilker05}. Lastly, and most speculative, is a mode that
creates clusters at the periphery of the \ion{H}{1} disk. We suggest that this is where gas that is accreted
joins the already existing gas and that this process leads to low level cluster formation. We consider these
radii to represent the outer banks of galaxies, where the waves of incoming gas break upon the shores
of the existing disk and leave telltale `foam' upon their arrival.

\acknowledgments{
This work relied on the
invaluable idlutils software library developed by M. R. Blanton, S.
Burles, D. P. Finkbeiner, D. W. Hogg, and D. J. Schlegel, and on the
Goddard IDL library maintained by W. Landsman.
We thank the THINGS and WHISP teams that 
produced the \ion{H}{1} maps used here. We thank the referee for suggestions that clarified the text.
DZ and SHF were partially supported under 
NASA LTSA NNG05GE82G and NSF AST-0307482.  
DZ thanks Cambridge University and New York University for their
hospitality during the final stages of this work.
}
\clearpage

\onecolumn

\begin{deluxetable}{lrrrrrrcrcccl}
\tablewidth{40.5pc}
\tablecaption{Sample Galaxies \label{tab:gals}}
\tablehead{\colhead{Name} &\colhead{$i$} &\colhead{PA} &\colhead{D} &\colhead{R25} &\colhead{$U_{date}$} &\colhead{$U_{exp}$} &\colhead{$V_{date}$} &\colhead{$V_{exp}$} &\colhead{$V_{lim}$} &\colhead{$V_{acor}$} &\colhead{$\sigma_{V_{acor}}$} &\colhead{Ref.} \\
&[deg]&[deg]&[Mpc]&[kpc]&&[sec]&&[sec]}
\startdata
IC 4182		&	23	&	90	&	4.7	&	4.1	&	5/11/07	&	1640 	&	5/11/07	&	1640 	&	27.0	&	$-$0.11	&	0.05	&	1		\\
NGC 3351	&	40	&	13	&	10.1	&	10.8	&	3/21/07 	&	1640		&	3/21/07	&	1640		&	27.0	&	$-$0.15	&	0.05	&	4,5,6,7	\\	
NGC 4736	&	8	&	122	&	4.7	&	7.6	&	2/21/07	&	1640 	&	2/21/07	&	1476 	&	27.5 	&	$-$0.13	&	0.03	&	2,3		\\
NGC 4826	&	61	&	115	&	7.5	&	10.2	&	2/10/08	&	1476 	&	4/24/07	&	1476 	&	26.0	&	$-$0.69	&	0.04	&	1,8,9		\\
NGC 5474	&	26	&	132	&	7.2	&	5.0	&	5/10/07	&	328   	&	5/10/07	&	1804 	&	26.0	&	$-$0.26	&	0.01 &	1,10		\\
NGC 6503	&	74	&	121	&	5.3	&	4.5	&	4/23/07	&	1312 	&	4/11,23/07	&	3280 	&	26.0	&	$-$0.46	&	0.01	&	1,11,12	\\
\enddata
\tablerefs{
1: \cite{Karachentsev04}; 
2: \cite{Karachentsev05}; 
3: \cite{Trujillo09}; 
4: \cite{Graham97}; 
5: \cite{Rubin75}; 
6: \cite{Buta88}; 
7: \cite{Swartz06}; 
8: \cite{Tonry01}; 
9: \cite{Nilson73}; 
10: \cite{Rownd94};
11: \cite{Makarova99};
12: \cite{Begeman87}}
\end{deluxetable}

\begin{deluxetable}{lrrrrrr}
\tablewidth{0pt}
\tablecaption{Sources in Color-Magnitude and Background-Subtracted Hess Diagrams \label{tab:CMD}}
\tablehead{\colhead{Name} &\colhead{Annulus} &\colhead{Area} &\colhead{$N_{CMD}$} &\colhead{$N_{Hess}$} &\colhead{Fraction} 
&\colhead{\ \ $\mu_{eff}$}\\
&[$R_{25}]$&[arcmin$^{-2}$]&&&[\%]&[mag arcsec$^{-2}$]}
\startdata
IC 4182		&	$1.0-1.5$	&	30.95	&	1950 	&	248 		&	   13 		&	29.62 	\\
		&	$1.5-2.0$	&	41.66	&	2244 	&	 $-$48 		&	 ... 		&	30.44 	\\
		&	$2.0-2.5$	&	43.91	&	2408 	&	 $-$7 		&	 ... 		&	29.98 	\\
\hline		
NGC 3351	&	$1.0-1.5$	&	40.22	&	2037   	&	448 		&	22 		&	$>32.44$	\\
	&	$1.5-2.0$	&	38.22	&	1593 	&	  83 		&	  5 		&	$>34.91$	\\
	&	$2.0-2.5$	&	37.52	&	1554 	&	  72 		&	  5 		&	$>31.12$  \\
\hline
NGC 4736	&	$1.0-1.5$	&	85.38	&	8517 	&	2774 	&	33 		&	29.46 	\\
&	$1.5-2.0$	&    104.88	&	7570 	&	516 		&	  7 		&	30.43 	\\
&	$2.0-2.4$	&      83.93	&	6135 	&	489 		&	  8 		&	$>30.99$	\\
\hline
NGC 4826	&	$1.0-1.5$	&	40.42	&	1445 	&	$-$33 		&	...		&	30.00 	\\
	&	$1.5-2.0$	&	52.90	&	1964 	&	30 		&	2 		&	33.11 	\\
	&	$2.0-2.5$	&	72.15	&	2670 	&	33 		&	1 		&	31.34 	\\
\hline
NGC 5474	&	$1.0-1.5$	&	19.96	&	699		&	427		&	61		&	28.85   	\\
	&	$1.5-2.0$	&	27.95	&	433		&	52		&	12		&	$>33.57$	  \\
	&	$2.0-2.5$	&	35.93	&	465		&	$-$25		&	...		&	32.14 	  \\
\hline
NGC 6503	&	$1.0-1.5$	&	 8.69 	&	  622 	&	369  		&	59 		&	28.36 	\\
	&	$1.5-2.0$	&	 9.94		&	  554 	&	265 		&	48 		&	28.94 	\\
	&	$2.0-2.5$	&     13.63		&	  511 	&	114 		&	22 		&	29.75 	\\
\enddata
\tablenotetext{\ }{Fraction is undefined
when $N_{Hess}$ is negative (there is no excess of disk sources over the background).  
 $\mu_{eff}$ is given as a lower limit when the total flux in the Hess diagram is negative.  
 Note that there exist cases where $N_{Hess}$ is negative but the total flux is not.  This depends on the distribution of signal across the entire Hess diagram.
  }\end{deluxetable}

\begin{deluxetable}{lrrrr}
\tablewidth{0pt}
\tablecaption{Sources in LBT-matched GALEX catalogs \label{tab:GALEX}}
\tablehead{\colhead{Name} &\colhead{Annulus} &\colhead{$N_{GALEX}$} &\colhead{$N_{Hess}$} &\colhead{Fraction} \\
&[$R_{25}]$&&&[\%]}
\startdata
IC 4182		&	$1.0-1.5$	&	5	&	348	&	1	 \\
IC 4182		&	$1.5-2.0$	&	28	&	69 	&	41	 \\
\hline
NGC 3351	&	$1.0-1.5$	&	12	&	548	&	2	 \\
NGC 3351	&	$1.5-2.0$	&	27	&	266  &	10	\\
\hline
NGC 4736	&	$1.0-1.5$	&	56	&	2738 &	2	\\
NGC 4736	&	$1.5-2.0$	&	172	&	704  &	24 	\\
\hline
NGC 5474	&	$1.0-1.5$	&	8	&	427  & 	2	\\ 
NGC 5474	&	$1.5-2.0$	&	16	&	52   &	31	\\
\enddata
\tablenotetext{\ }{Because the GALEX catalogs already had so few usable sources 
we did not mask the regions we did in our LBT mosaics around bright stars; 
therefore, so that the fields are of similar area, the $N_{Hess}$ listed here are the values 
from our LBT catalogs before the bright star masks were applied. }\end{deluxetable}


\clearpage

}

\clearpage

\end{document}